\documentclass[twocolumn,secnumarabic,amssymb, nobibnotes, aps, prb, superscriptaddress]{revtex4-2}
\usepackage{lipsum}
\usepackage{titlesec}
\usepackage{amsmath,amssymb,xcolor}
\titlespacing\section{0pt}{12pt plus 4pt minus 4pt}{1pt plus 20pt minus 2pt}
\usepackage{xcolor}
\usepackage{amsmath}
\usepackage{comment}
\usepackage{physics}
\usepackage[colorlinks,bookmarks=false,citecolor=red,linkcolor=blue,urlcolor=blue]{hyperref}
\usepackage{graphicx,epstopdf}
\usepackage{hyperref}
\usepackage{ulem}
\hypersetup{
    colorlinks,
    citecolor=blue,
    filecolor=blue,
    linkcolor=blue,
    urlcolor=blue} 
\usepackage{array,multirow}
\newcolumntype{C}[1]{>{\centering\arraybackslash}m{#1}}
\usepackage{afterpage}
\usepackage{placeins}
\usepackage{graphicx}
\usepackage{float}
\usepackage{booktabs}
\usepackage{multirow}
\usepackage{array}
\usepackage{setspace}
\graphicspath{{Figs/}}
\usepackage{siunitx}
\usepackage{hhline}
\usepackage{xfrac}
\usepackage{float,graphicx}
\usepackage{mathtools}
\usepackage{listings}
\usepackage{amssymb}
\usepackage{titlesec}
\usepackage{amsfonts}
\usepackage[version=4]{mhchem}

\usepackage{epstopdf}
\catcode`@11
\def\seceqaa{\@addtoreset{equation}{section}
\def\theequation{A\arabic{equation}}}
\def\seceqbb{\@addtoreset{equation}{section}
\def\theequation{B\arabic{equation}}}
\def\seceqcc{\@addtoreset{equation}{section}
\def\theequation{C\arabic{equation}}}
\def\seceqdd{\@addtoreset{equation}{section}
\def\theequation{D\arabic{equation}}}
\def\seceqee{\@addtoreset{equation}{section}
\def\theequation{E\arabic{equation}}}
\def\seceqff{\@addtoreset{equation}{section}
\def\theequation{F\arabic{equation}}}
\def\seceqgg{\@addtoreset{equation}{section}
\def\theequation{G\arabic{equation}}}
\def\seceqhh{\@addtoreset{equation}{section}
\def\theequation{H\arabic{equation}}}
\catcode`@11
\begin{document}
\title{Spin-1/2 string correlations and singlet-triplet gaps of frustrated ladders with  ferromagnetic (F) legs and alternate F and AF rungs}

\author{Monalisa Chatterjee}
\affiliation{S. N. Bose National Centre for Basic Sciences, Block-JD, Sector-III, Salt Lake, Kolkata 700106, India}

\author{Manoranjan Kumar}
\email{manoranjan.kumar@bose.res.in}
\affiliation{S. N. Bose National Centre for Basic Sciences, Block-JD, Sector-III, Salt Lake, Kolkata 700106, India}

\author{Zolt\'an G. Soos}
\email{soos@princeton.edu}
\affiliation{Department of Chemistry, Princeton University, Princeton, New Jersey 08544, USA}

\begin{abstract}
The frustrated ladder with alternate ferromagnetic(F) exchange $–J_F$ and AF exchange $J_A$ to first neighbors and F exchange $–J_L$ to second neighbors is studied by exact diagonalization (ED) and density matrix renormalization group (DMRG) calculations in systems of $2N$ spins-1/2 with periodic boundary conditions. The ground state is a singlet $(S = 0)$ and the singlet-triplet gap $\varepsilon_T$ is finite for the exchanges considered. Spin-1/2 string correlation functions $g_1(N)$ and $g_2(N)$ are defined for an even number $N$ of consecutive spins in systems with two spins per unit cell; the ladder has string order $g_2(\infty)> 0$ and $g_1(\infty) = 0$. The minimum $N^*$ of
$g_2(N)$ is related to the range of ground-state spin correlations. Convergence to $g_2(\infty)$ is from below, and $g_1(N)$ decreases exponentially for $N \geq N^*$. Singlet valence bond (VB) diagrams account for the size dependencies. The frustrated ladder at special values of $J_F$, $J_L$ and $J_A$ reduces to well-known models such as the spin-1 Heisenberg antiferromagnet and the $J_1-J_2$ model, among others. Numerical analysis of ladders matches previous results for spin-1 gaps or string correlation functions and extends them to spin-1/2 systems. The nondegenerate singlet ground state of ladder is a bond-order wave, a Kekul\'e VB diagrams at $J_L = J_F/2 \leq J_A$, that is reversed on interchanging $–J_F$ and $J_A$. Inversion symmetry is spontaneously broken in the dimer phase of the $J_1-J_2$ model where the Kekul\'e diagrams are the doubly degenerate ground states at $J_2/J_1 = 1/2$.    
\end{abstract}
\pacs{}
\maketitle
\section{Introduction}
The spin-1/2 Heisenberg antiferromagnet (HAF) with isotropic AF exchange $J_1 > 0$
between first neighbors has been central to theoretical studies of correlated many-spin systems,
including the famous exact $1D$ solution based on the Bethe ansatz \cite{Bethe1931} and the magnetism of
inorganic \cite{Jongh1974} and organic \cite{Deiseroth1983} materials that contain $1D$ spin-1/2 chains.
The addition of AF exchange $J_2 > 0$ between second neighbors introduces frustration and leads to
interesting ground state properties such a bond order waves \cite{Kumar2012}, spiral phases \cite{SR_White_1996,Kumar2010,Kumar2015} and spin liquids \cite{Savary2017} due to
quantum fluctuations. The $J_1 -J_2$ model has been successfully applied to the magnetism of crystals
with $1D$ chains of $S = 1/2$ of transition metal ions such as $Cu(II)$ \cite{Hase1993,Sudip2020}.

\begin{figure}[h!]
\includegraphics[width=\linewidth]{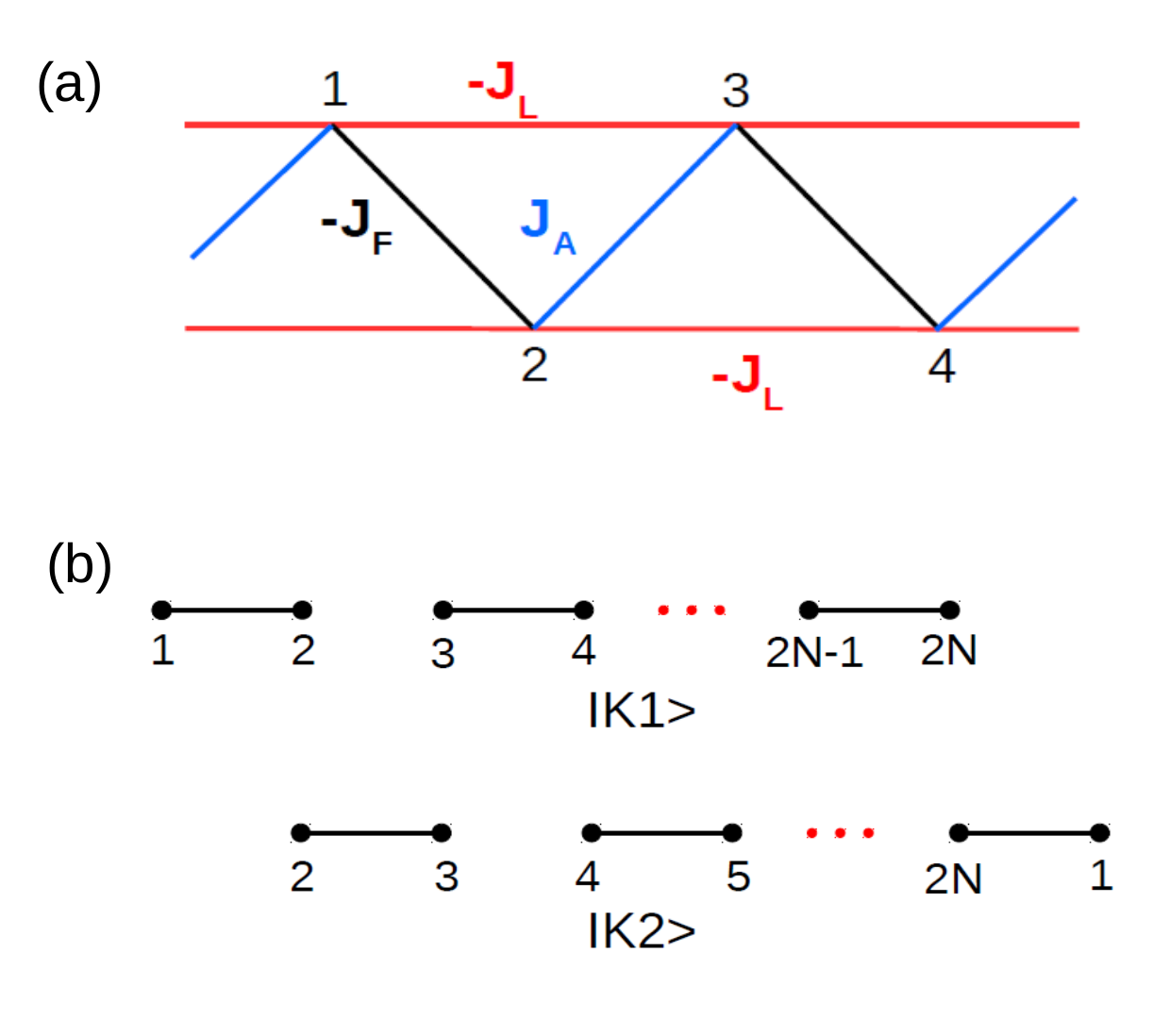}
\caption{(a) The F-AF spin-1/2 ladder with F exchange $-J_L < 0$ between spins $r$ and $r + 2$ in either leg, F exchange $-J_F$ in rungs $2r - 1$, $2r$ and AF exchange $J_A$ in rungs $2r$, $2r + 1$. (b) Kekul\'e diagrams $\lvert K1\rangle$ and $\lvert K2\rangle$ with singlet-paired spins $(2r – 1,2r)$ and $(2r,2r + 1)$, $r = 1$ to $N$.}
\label{fig1_p}
\end{figure}

Dimerized chains have lower symmetry and different $J_1$ with neighbors to the right and left. The AFAF model \cite{Hida1992} has alternate $J_A = 1$ to one neighbor and variable $-J_F$ to the other. The model has
attracted much attention since its approximate realization in some materials, eg. $Na_2Cu_2TeO_6$ \cite{Lin2022, Gao2020}, $CuNb_2O_6$ \cite{Kodama1999} and $(CH_3)_2NH_2CuCl_3$ \cite{Stone2007}.
The AFAF model is the frustrated F-AF ladder in Fig. \ref{fig1_p} with spin $S_r= 1/2$ at site $r$ and $J_L = 0$. The recent study \cite{Tseng2023} of weakly-doped $Sr_{14}Cu_{24}O_{41}$ using resonant inelastic X-ray scattering illustrates the scope spin-1/2 ladders. Other two leg ladders singlet ground states may exhibit  superconductivity on tuning the exchange interactions.  \cite{Dagotto1992,Patel2016,Zhang2017}. 

The $J_F \to \infty$ limit of the AFAF model is the spin-1 HAF that has been intensively studied theoretically and numerically since Haldane predicted it to be gapped \cite{Haldane1983}. The ground state of the AFAF model with exchange $-J_L$ between second neighbors has interesting topological
properties \cite {Hida2013,Sahoo2020} as do AFAF models \cite{Sahoo2020} with spins $ S>1/2$, that are the focus of current research. The topological properties of AFAF models with $J_2 < 0$ pose open problems.

We study in this paper the F-AF ladder in Fig. \ref{fig1_p} with three isotropic exchanges: F
exchange $-J_L$ between neighbors $r$, $r + 2$ in legs, F exchange $-J_F$ at rungs $2r - 1$, $2r$ and AF exchange $J_A$ at rungs $2r$, $2r + 1$. We consider parameters $J_L$ , $J_F$ and $J_A = 1$ leading to a singlet $(S = 0)$ ground state $G(J_L, J_F)$. The ladder reduces to important models in special cases. It is the spin-1/2 HAF at $J_L = 0$ and $-J_F = J_A$ with one spin per unit cell and the AFAF model at $J_L = 0$ and $-J_F \neq J_A$ with two spins per unit cell. The ladder is frustrated except when $J_L = 0$ or $J_F = 0$. The limit $J_F \to \infty$ is the spin-1 HAF with $J = (1 - 2J_L )/4 > 0$ between adjacent F rungs. The limit $J_L \to \infty$ is a $J_1 -J_2$ model with $J_1 = (1-J_F )/2 > 0$ and $J_2 = -J_L$ . The symmetry is higher \cite{chatterjee2023} at infinite $J_F$ or $J_L$ .

The spin Hamiltonian with $J_A = 1$ as the unit of energy is
\begin{eqnarray}
\label{ssm_eq1}
{H_{F-AF}(J_L,J_F)}=\sum_{r=1}^{N} (\vec{S}_{2r} \cdot \vec{S}_{2r+1}-J_F \vec{S}_{2r-1}  \cdot \vec{S}_{2r})- \nonumber \\ 
& & \hspace*{-160pt} J_L\sum_{r=1}^{2N} \vec{S}_r \cdot \vec{S}_{r+2}.
\end{eqnarray}
The total spin $S \leq N$ and its z component $S^z$ are conserved. We consider systems of $2N$ spins
with periodic boundary conditions and seek the thermodynamic limit $N \to \infty$. The ground state
$G(J_L, J_F)$ in that limit has two noteworthy features. First, it is either a singlet or ferromagnetic \cite{Dmitriev2000} for any $J_L, J_F$ and $J_A = 1$. Second, the exact $G(J_L, J_F)$ is a product of singlet-paired spins along a line where F exchanges cancel exactly. Both are central in the following. A product of singlet-paired spins is the exact ground state at special points of other $1D$ and $2D$ spin-1/2 systems \cite{Majumder1969,SRIRAMSHASTRY1981,Sahoo2020,Furukawa2011,White1996,Brijesh2022,Hong2013,Tsirlin2008,Biswal2023,Chitra1995,Tassel2010}.

We develop three themes. The first is string correlation functions in spin-1/2 chains. Den Nijs and Rommelse \cite{Nijs1989} and Tasaki \cite{Tasaki1992} pointed out a hidden $Z_2 \times Z_2$ symmetry that can be measured by string correlation functions. Oshikawa \cite{Oshikawa1992} generalized the symmetry to Haldane chains with arbitrary integer $S > 1$. The critical theory of quantum spin chains by Affleck and Haldane \cite{AffleckHaldane} includes models with half-integer $S$ and $Z_2$ symmetry. All the models considered \cite{Oshikawa1992,AffleckHaldane} have equal isotropic exchange between either integer or half-integer S. The F-AF ladder has instead alternate $-J_F$ and $J_A = 1$ between first neighbors. It has two spins per unit cell in general, two string correlation functions and $Z_2$ symmetry only in limits with one spin per unit cell. 

The string correlation function $O(p-p^\prime)$ between consecutive spins from $p$ to $p^\prime$ in the spin-$1$ HAF is finite in the limit $\lvert p - p^\prime\rvert \to \infty$. Hida \cite{Hida1992} adapted the spin-$1$ expression to string correlation functions of the AFAF model ($J_L = 0$ in Eq. \ref{ssm_eq1}) with open boundary conditions. The string correlation functions $O(r - r^\prime)$ necessarily have an even number of consecutive spins-$1/2$ in Eq. \ref{ssm_eq2} of ref. \cite{Hida1992}.


We use this expression in general. The string correlation function for an even number $N$ of consecutive spins-1/2 is
\begin{eqnarray}
\label{ssm_eq2}
{g_{1}(N)}=\langle G\rvert exp (i\pi \sum_{j=1}^{N} {S}_j^{z}) \lvert G \rangle.
\end{eqnarray}
The expectation value is with respect to the ground state in the thermodynamic limit or in finite systems with periodic boundary conditions. The general expression for spin-1/2 strings is well defined without reference to the spin-1 HAF. The initial spin is arbitrary in systems with one spin per unit cell. Since the F-AF ladder has two, the string correlation function $g_2(N)$ runs from $j = 2$ to $N + 1$ in Eq. \ref{ssm_eq2}. In either case string correlation functions of $2p \leq 2N$ spins can be evaluated for $2N$-spin ladders. The choice $2p = N$ is convenient for taking the thermodynamic limit.

The exact ground state along the line $J_L = J_F/2 \leq 1$ is the Kekul\'e valence bond (VB) diagram $\lvert K2\rangle$ in Fig. \ref{fig1_p} with singlet-paired spins $2r$, $2r + 1$ shown as lines, and as shown in Sec. \ref{sec-IV}, $g_2(N) = 1$, $g_1(N) = 0$ at any system size. To evaluate string correlation functions, we obtain the ground state $G(J_L,J_F,2N)$ in increasingly large systems of $2N$ spins using exact diagonalization (ED) and density matrix renormalization group (DMRG) calculations. We interpret the results in terms of VB diagrams.

VB diagrams are an explicit general way to construct \cite{Soos1984,Ramasesha1984} correlated many-spin states in real space with conserved $S \leq N$ for $2N$ spins-1/2. The spins are placed at the vertices of the regular $2N$ polygon. A line $(m,n)$ between vertices $m$ and $n$ represents normalized singlet-paired spins whose phase is fixed by $m < n$
\begin{eqnarray}
\label{ssm_eq3}
{(m,n)}=(\alpha_m \beta_n-\beta_m \alpha_n)/\sqrt{2}.
\end{eqnarray}
A legal (linearly independent) singlet diagram $\lvert q\rangle$ has $N$ lines $(m,n)$, an $N$-fold product of singlet-paired spins, that connects all $2N$ vertices once without any crossing lines. Diagrams with crossing lines are not linearly independent since they can be resolved into legal diagrams. The normalized singlet ground state is formally a linear combination of singlet diagrams,
\begin{eqnarray}
\label{ssm_eq4}
{\lvert G(J_L,J_F,2N)\rangle}=\sum_{q}C(q,J_L,J_F)\lvert q\rangle.
\end{eqnarray}
The sum is over $R_0(2N)$ singlet diagrams that depends only on system size. The coefficients $C(q,J_L,J_F)$ depend on models, parameters and boundary conditions as well as system size. We find below the diagrams $\lvert q\rangle$ that are eigenfunctions of the string operator in Eq. \ref{ssm_eq2}. The VB analysis accounts for the remarkable result of increasing string correlation functions $g_2(N)$ with system size. Convergence to string order $g_2(\infty)$ is from below.

The second theme is to recognize three regimes of the F-AF ladder in the positive quadrant of the $J_L$, $J_F$ plane. Near the origin, in Eq. \ref{ssm_eq1} is a system of $N$ dimers with exchange $J_A = 1$ between spins $2r$, $2r + 1$, a singlet ground state, and frustrated F interactions between adjacent dimers. The singlet-triplet gap $\varepsilon_T(J_L,J_F)$ is large, spin correlations are short ranged, and small systems suffice for the thermodynamic limit. Increasing $J_F > 1$ while maintaining a singlet ground state leads to $N$ rungs $2r-1$, $2r$ with triplet $(S = 1)$ ground states and net AF exchange $1 - 2J_L > 1$ between adjacent rungs. Increasing $J_L > 1$ while maintaining a singlet ground state leads to F legs with net AF exchange $1 - J_F > 0$ between spins in different legs. Results for general $J_L$, $J_F$ are understood qualitatively this way.

The third theme is dimerization. The nondegenerate singlet ground state of Eq. \ref{ssm_eq1} is a bond order wave (BOW). The bond orders along the line $J_L = J_F/2 \leq 1$ are $\langle S_2 \cdot S_3\rangle = -3/4$ for singlet-paired spins and $\langle S_1 \cdot S_2\rangle = 0$ due to cancelling F exchanges. Interchanging $-J_F$ and $J_A = 1$ reverses the BOW without changing the energy spectrum. Increasing $J_F$ reduces the BOW to \cite{chatterjee2023} $\langle S_1 \cdot S_2\rangle = 1/4$ and $\langle S_2 \cdot S_3\rangle = -0.350$ in the limit $J_F \to \infty$. Increasing $J_L$ to infinity leads to $\langle S_1 \cdot S_2\rangle =\langle S_2 \cdot S_3\rangle = -1/4$ and suppresses dimerization.

We discuss F-AF ladders with parameters $J_L$, $J_F$ in in Eq. \ref{ssm_eq1} leading to singlet ground states. The paper organized as follows. Sec. \ref{sec-II} summarizes the numerical methods used to obtain thermodynamic limits. Sec. \ref{sec-III} presents the singlet-triplet gap $\varepsilon_T(J_L,J_F)$ in the three regimes. String correlation functions $g_1(N)$ and $g_2(N)$ are defined in Sec. \ref{sec-IV} for spin-1/2 systems with two spins per unit cell. The $g_2(N)$ minimum at $N^*$ is a collective estimate of the range of ground-state spin correlations, while $g_1(N)$ decreases exponentially with system size for $N \geq N^*$. In Sec. \ref{sec-V} we consider string correlation functions of the $J_1-J_2$ model with one spin per unit cell and spontaneous dimerization for some parameters. Sec. \ref{sec-VI} is a brief summary.

\section{ Methods \label{sec-II}}
We use two numerical methods, ED and DMRG, to solve Eq. \ref{ssm_eq1} at $J_A = 1$ and variable $J_L$, $J_F$ in sectors with $S^z = 0$ or $1$ for $2N$ spins-1/2 or for the HAF with $n$ spins-1. ED up to $24$ spins-1/2 is sufficient for the thermodynamic limit of systems with short-range correlations or large $\varepsilon_T(J_L,J_F)$. DMRG with periodic boundary conditions is used for larger systems. The ground state is a singlet when the lowest energy in the $S^z= 0$ sector does not appear in other sectors. We also perform VB calculations to obtain the coefficients $C(q)$ in Eq. \ref{ssm_eq4} in systems of $2N \leq 16$ spins.

DMRG is a well-established numerical technique for the ground state and low-lying excited states of correlated $1D$ systems \cite{White1992,Schollwock2005,Hallberg2006}. We use a modified DMRG algorithm that adds four new sites (instead of two) to the superblock at each step \cite{Sudip2019}. This avoids interaction terms between old blocks in models with second neighbor exchange, here $-J_L$. All calculations are performed with periodic boundary conditions. We obtain truncation errors of $10^{-10}$ or less on keeping $512$ eigenvectors of the density matrix and $4$ or $5$ finite sweeps. Systems up to $2N = 192$ spins-1/2 or $n = 64$ spins-1 were used for finite-size scaling.

There are additional external and internal checks on the accuracy of DMRG calculations. External checks are spin-1 calculations using other numerical methods \cite{Tasaki1992,Nomura1989} or DMRG with open boundary conditions \cite{Huse1993}. Excellent agreement for spin correlation functions to $6$ or $7$ decimal places is due at least partly to the large Haldane gap \cite{Huse1993} of the spin-1 HAF. Internal checks rely on the singlet/F boundary \cite{Dmitriev2000} at
\begin{equation}
{J_F}= 2J_L/(2J_L-1),    \hspace{0.8cm} 2J_L,J_F\geq 1.
\label{ssm_eq5}
\end{equation}   
The F energy per dimer is independent of system size,
\begin{equation}
{\varepsilon_F}(J_L,J_F)= -(2J_L+J_F-1)/4.
\label{ssm_eq6}
\end{equation}
The singlet ground state per dimer, $\varepsilon_0(J_L,J_F,2N)$, is size dependent in general but must become size independent at the boundary. ED returns two states with $S^z = 0$ and one with $S^z = 1$ at $\varepsilon_0 = \varepsilon_F$. There are additional $S^z = 0$ and $1$ states just above $\varepsilon_F$. The DMRG accuracy at $2N = 32$ drops to $4$ or $5$ decimal places for the dense spectrum at the boundary.

\section{Singlet-triplet gap \label{sec-III}}

Near the origin of the $J_L$, $J_F$ plane, the singlet ground state energy per dimer is conveniently written in terms of $J_\pm = J_L \pm J_F/2$,
\begin{equation}
{\varepsilon_0}(J_L,J_F)=-\frac{3}{4}-\frac{3J_-^{2}}{4(2+J_+)}+ 0(J_-^3).
\label{ssm_eq7}
\end{equation}
Spins $2r$, $2r + 1$ are dimers with exchange $J_A = 1$ and $J_L$, $J_F$ 
cancel exactly when $J_- = 0$. The range of $J_-$ at constant $J_+$ 
is from $-J_+$ at $(0,J_F)$ to $J_+$ at $(2J_L,0)$. The ground state is a singlet for $J_+ \leq 2$, degenerate with $\varepsilon_F = -3/4$ at $J_L = 1$, $J_F = 2$. The virtual states at $(2 + J_+)$ in second-order perturbation theory are singlet linear combinations of adjacent triplet dimers. ED results for $\varepsilon_0(J_L,J_F,2N) + 3/4$ are listed in Table \ref{tab:table1} for $2N = 16$ and $24$ at constant $J_+$ and $J_-=\pm J_+$. The size dependence is weak. Differences at $J_-=\pm J_+$ are 
of order $J_-^{3}$.
\begin {table}[ht]
\normalsize
\caption{\label{tab:table1} 
Ground-state energy $\varepsilon_0$ per dimer at constant $J_+$ and $J_- = \pm J_+$. ED at system sizes $16$ and $24$; the $J_-^{2}$ term of Eq. \ref{ssm_eq7}.}
\begin{center}
\begin{tabular}{|C{1.5cm}|C{1.5cm}|C{1.4cm}|C{1.4cm}|C{1.6cm}|} 
\hline
$J_+=J_L+J_F/2$ & $J_-=J_L-J_F/2$ & $\varepsilon_0(16)+3/4$ & $\varepsilon_0(24)+3/4$ & $J_-^{2}$ term, Eq. \ref{ssm_eq7} \\ 
\hline
\hline
\multirow {2} {*} {$0.4$} & $0.4$ & $-0.0486$ & $-0.0486$ & \multirow {2} {*} {$-0.05$}\\
\cline{2-4}
 &  $-0.4$ & $-0.0498$ & $-0.0498$ & \\
\hline
\multirow {2} {*} {$0.8$} & $0.8$ & $-0.1604$ & $-0.1599$ & \multirow {2} {*} {$-0.1714$}\\
\cline{2-4}
 &  $-0.8$ & $-0.1679$ & $-0.1678$ & \\
 \hline
\multirow {2} {*} {$1.2$} & $1.2$ & $-0.3069$ & $-0.3046$ & \multirow {2} {*} {$-0.3375$} \\
\cline{2-4}
&  $-1.2$ & $-0.3220$ & $-0.3217$ & \\
\hline
\multirow {2} {*} {$1.6$} & $1.6$ & $-0.4742$ & $-0.4695$ & \multirow {2} {*} {$-0.5333$}\\
\cline{2-4}
&  $-1.6$ & $-0.4945$ & $-0.4939$ & \\
\hline
\end{tabular}
\end{center}
\end {table}

\begin{figure}[h!]
\includegraphics[width=\linewidth]{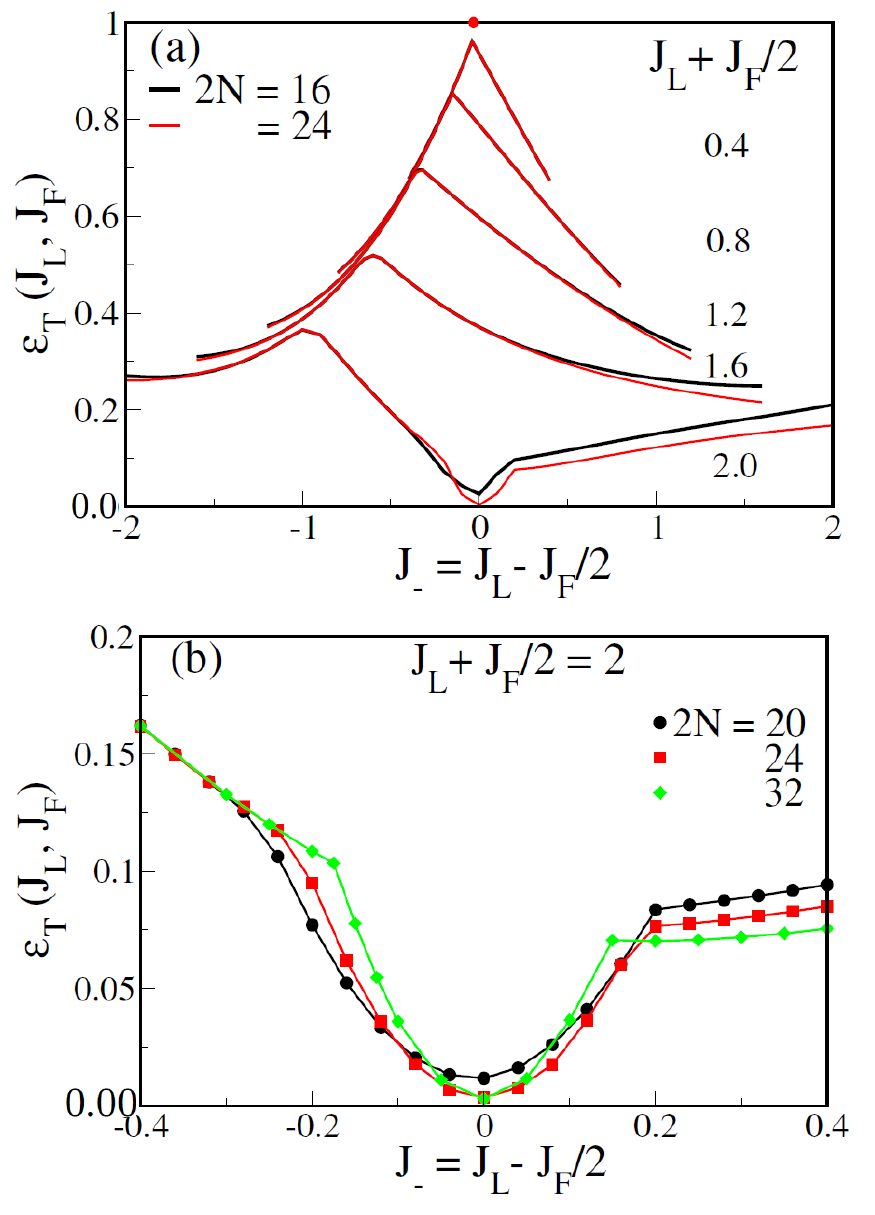}
\caption{(a) Singlet-triplet gap $\varepsilon_T(J_L,J_F)$ vs. $J_- = J_L - J_F/2$ at constant $J_+ = J_L + J_F/2$ and system sizes $2N = 16$ and $24$. The range is $-J_+ \leq J_- \leq J_+$, and $\varepsilon_T(0,0) = 1$. (b) $\varepsilon_T(J_L,J_F)$ at constant $J_+ = 2$ vs. $-0.4 \leq J_- \leq 0.4$ at $2N = 20$, $24$ and $32$. The crossing points are $\varepsilon_T = N(\varepsilon_F - \varepsilon_0)$.}
\label{fig2_p}
\end{figure}

The upper panel of Fig. \ref{fig2_p}, shows $\varepsilon_T(J_L,J_F)$ at constant $J_+$ for $2N = 16$ and $24$ as a 
function of $-J_+\leq J_-\leq J_+$. The gap decreases from $\varepsilon_T(0,0) = 1$ with increasing $J_+$ and is asymmetric in $J_-$. The size dependence is weak except at $J_+ = 2$, $J_-> 0$. The cusp at $J_- \approx 0$ and $J_+ = 0.4$ is due to lifting the $N$-fold degeneracy of localized triplets at $2r$, $2r + 1$. The lowest triplet is nondegenerate with wavevector $k = 0$ or $\pi$ that switches from $\pi$ to $0$ with increasing $J_-$. 

The lower panel of Fig. \ref{fig2_p} zooms in on $\varepsilon_T(J_L,J_F,2N)$ at $J_+ = 2.0$ and $–0.4 \leq J_-\leq 0.4$, which includes the singlet/F boundary at $J_- = 0$. The crossing points at positive and negative $J_-$ are due to finite size. The singlet and F ground states are extensive while $\varepsilon_T$ is intensive. In finite systems, the extensive difference $N(\varepsilon_F - \varepsilon_0)$ at the singlet/F boundary is a parabola, $-N$ times the $J_-^{2}$ term of Eq. \ref{ssm_eq7}. The calculated $\varepsilon_T(J_L,J_F,2N)$ at $J_+=2$, $J_-=0$ are $\varepsilon_T (20)=0.0118$, $\varepsilon_T(24)=0.0037$ and $\varepsilon_T(32) \approx 0.0031 \pm 0.001$. As mentioned in Sec. \ref{sec-II}, the dense spectrum at the boundary limits the numerical accuracy. The gaps of finite ladders in the lower panel are well approximated by parabolas with finite $\varepsilon_T \approx 0.003$ at $J_- = 0$, crossing points at $\varepsilon_T = N(\varepsilon_F – \varepsilon_0)$ and asymmetry due to $J_-^{3}$. 

\begin{figure}[h!]
\includegraphics[width=\linewidth]{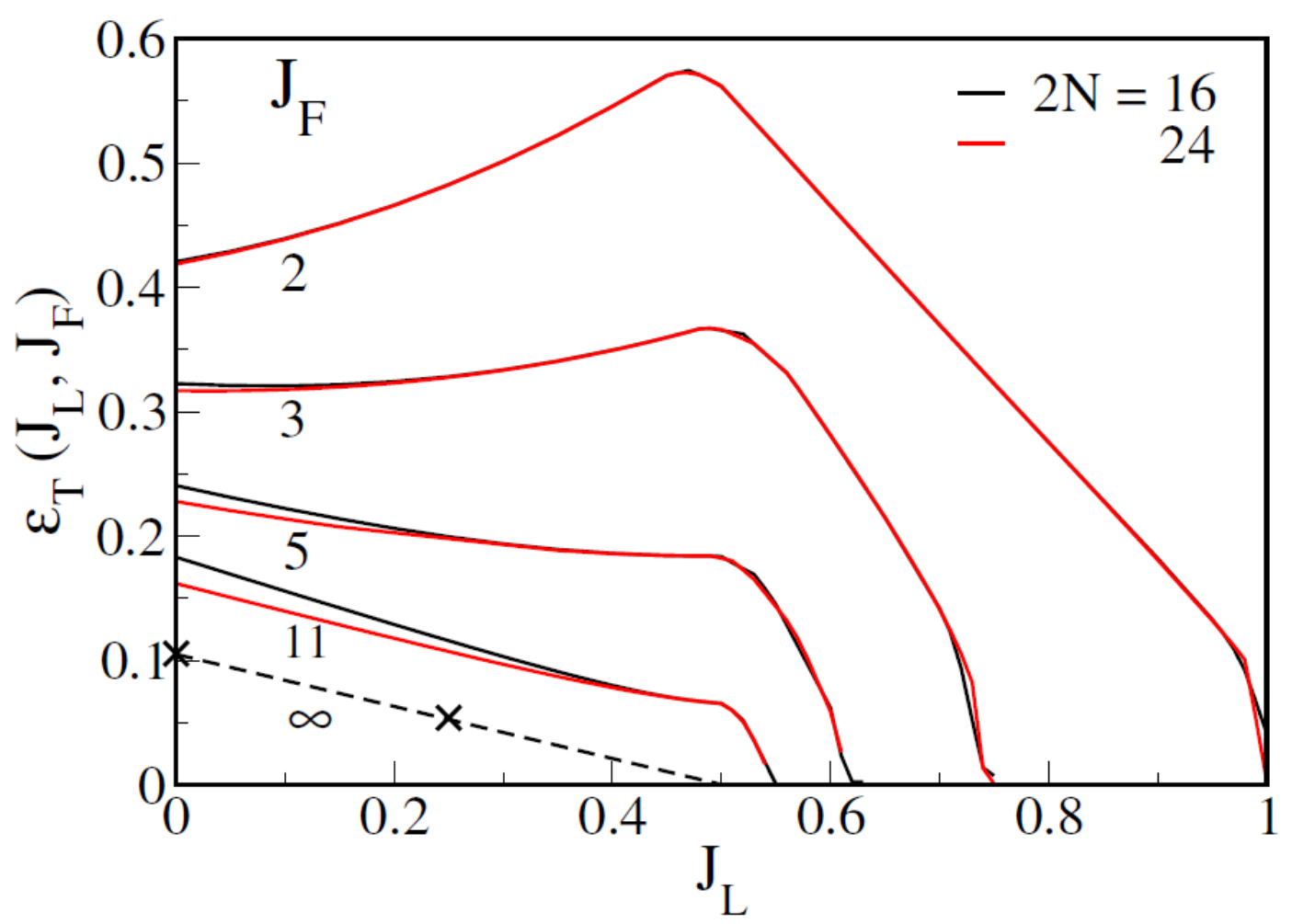}
\caption{Singlet-triplet gap $\varepsilon_T(J_L,J_F)$ vs. $J_L$ at constant $J_F$ and system sizes $2N = 16$ and $24$. The gaps are $< 0.005$ at the singlet/F boundary, $2J_L = J_F/(J_F - 1)$. The dashed line is $(1 - 2J_L)\Delta(1)/4$ where $\Delta(1) = 0.4105$ is the Haldane gap;\cite{Huse1993} the crosses at $J_L = 0$ and $0.25$ are for $J_F = 200$ and $2N = 64$.
}
\label{fig3_p}
\end{figure}

The size dependence of $\varepsilon_T(J_L,J_F)$ is much weaker at $J_F > 2$ than at $J_L > 1$. In either case the singlet/F boundary, Eq. \ref{ssm_eq5}, limits the magnitude of the other exchange. Fig. \ref{fig3_p} shows $\varepsilon_T(J_L,J_F)$ at system sizes $2N = 16$ and $24$ as a function of $J_L$ at the indicated $J_F$. The maximum gap decreases and broadens with increasing $J_F$ where triplets with different wave vectors are closely spaced. The $k = 0$ triplet is lowest when the gap is decreasing and (almost) vanishes at the singlet/F boundary. The $k = \pi$ triplets at $J_L = 0$ have the strongest size dependence.

The spins $2r - 1$, $2r$ form triplets when $J_F$ is large; the ground state degeneracy is $3^N$ at $J_A = J_L = 0$. To study the large $J_F$ regime of the ladder, we rewrite Eq. \ref{ssm_eq1} as
\begin{eqnarray}
\label{ssm_eq8}
{H_{F-AF}(J_L,J_F)}= -J_F\sum_{r=1}^{N}  \vec{S}_{2r-1} \cdot \vec{S}_{2r}+ (1-2J_L)/4 \nonumber \\ 
&&\hspace*{-205pt} \times \sum_{r=1}^{N} (\vec{S}_{2r-1}+\vec{S}_{2r})\cdot (\vec{S}_{2r+1}+
\vec{S}_{2r+2})+V'.
\end{eqnarray}
The first term corresponds to noninteracting dimers with triplet ground states. The second term is the spin-1 HAF with exchange $J = (1 - 2J_L)/4 > 0$ between neighboring rungs $2r,2r - 1$. The operator $V^\prime$ contains all other exchanges. The coefficients are: $(3 + 2J_L)/4$ for exchange between spins $2r$, $2r + 1$; $-(1 + 2J_L)/4$ for exchange between spins $r$ and $r + 2$; and $-(1 - 2J_L)/4$ for exchange between spins $2r -1$ and $2r + 3$. Virtual excitations at finite $J_F$ lead to effective Hamiltonians with excitations of order $1/J_F$. Eq. \ref{ssm_eq8} adiabatically connects ladders with finite $J_F$ and $2J_L < 1$ to the spin-1 HAF with $V^\prime = 0$ in the limit $J_F \to \infty$.

DMRG with open boundary conditions returns \cite{Huse1993} $\Delta(1) = 0.4105$ for the Haldane gap.
We find $\Delta(1) = 0.4106$ for $48$ spins-1 and periodic boundary conditions. The dashed line in Fig. \ref{fig3_p} is $(1- 2J_L)\Delta(1)/4$. The gaps indicated by crosses at $J_L = 0$ and $1/2$ are ${0.1055}$ and ${0.0538}$, respectively, at $J_F = 200$ and system size $2N = 64$. 

\begin{figure}[h!]
\includegraphics[width=\linewidth]{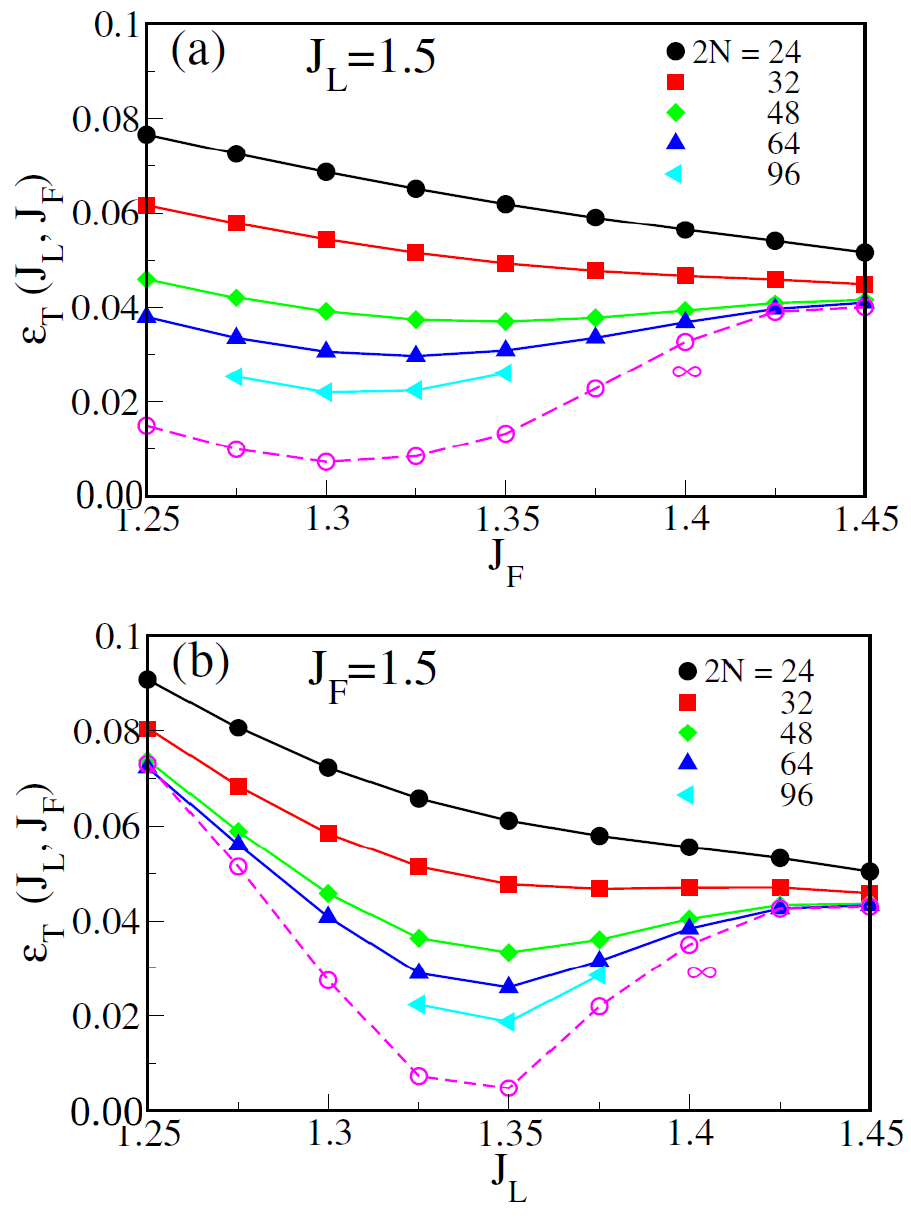}
\caption{Size dependence of singlet-triplet gaps $\varepsilon_T(J_L,J_F)$ at (a) constant $J_L = 1.5$, variable $J_F$ and (b) constant $J_F = 1.5$, variable $J_L$. The dashed line is the $1/N$ extrapolation from $2N = 24$ to $2N = 64$ or $96$.}
\label{fig4_p}
\end{figure}

The size dependence of $\varepsilon_T(J_L,J_F)$ is shown in Fig. \ref{fig4_p} at constant $J_L = 1.5$, variable $J_F$ in the upper panel and at constant $J_F = 1.5$, variable $J_L$ in the lower panel. The singlet/F boundary is $(1.5,1.5)$ where the estimated gap is $< 0.005$. Both panels show $\varepsilon_T(J_L,J_F,2N)$ minima in finite ladders and increasing gaps with weak size dependence at $(1.5,1.45)$ or $(1.45,1.5)$. The dashed lines are $1/N$ extrapolations. The remarkably small gap in Fig. 4(a) has been noted \cite{Hida2013} previously using DMRG with open boundary conditions. Small $\varepsilon_T(J_L,J_F)$ with a minimum are found at $J_A = 1$ and comparable $J_L$, $J_F$ with $J_L + J_F \approx 2.8$. We do not have an explanation for a minimum gap.

\begin{figure}[h!]
\includegraphics[width=\linewidth]{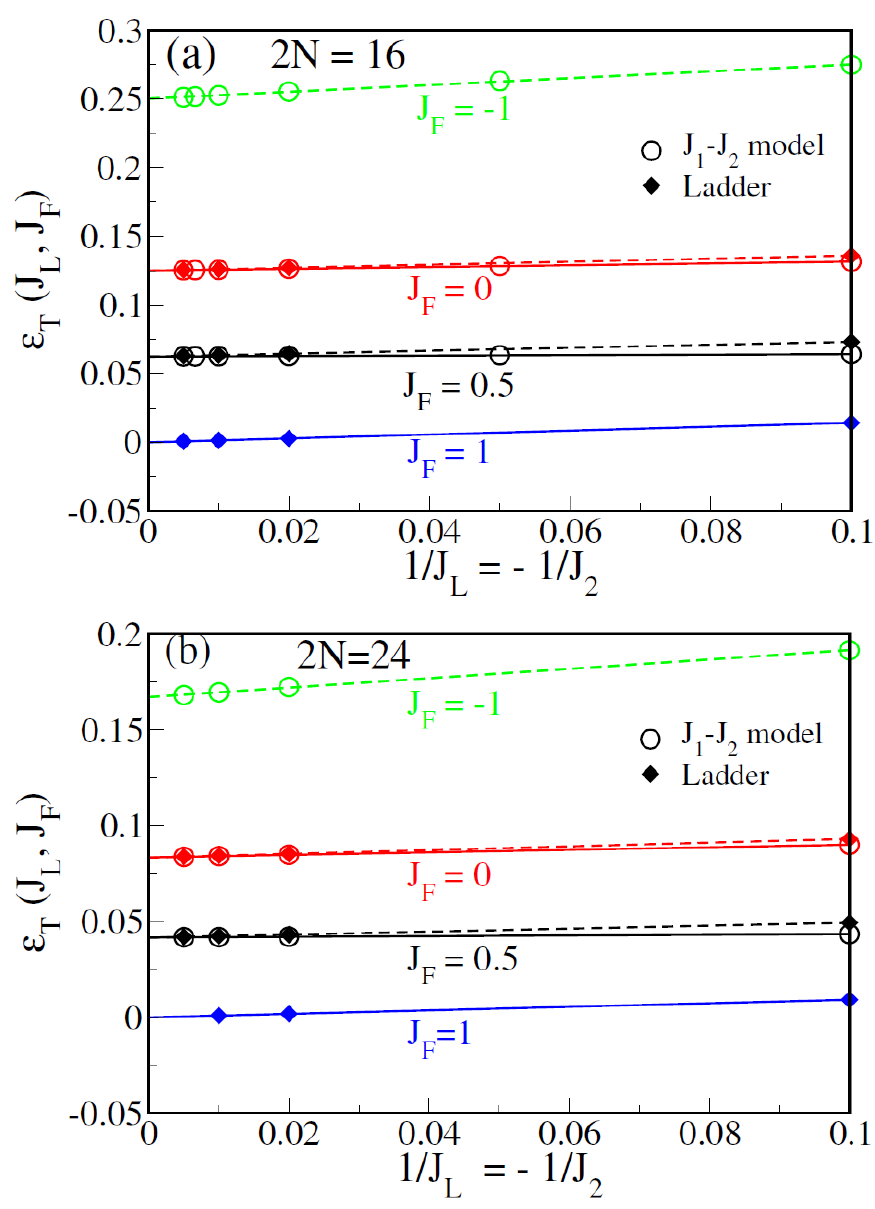}
\caption{Singlet-triplet gaps $\varepsilon_T(J_L,J_F)$ vs. $1/J_L = -1/J_2$ at constant $J_F = -1, 0, 0.5$ and $1$ at system sizes $2N = 16$ in (a) and $24$ in (b). Open symbols refer to $J_1-J_2$ models with $J_1 = (1 - J_F)/2$ and $J_2 = -J_F$. The ladder at $J_F = -1$ is a $J_1-J_2$ model. The $1/J_L = 0$ gaps are 
$(1 - J_F)/2N$ for both. The $J_F = 1$ gap is finite in ladders, zero in $J_1-J_2$ models.}
\label{fig5_p}
\end{figure}

Fig. \ref{fig5_p} shows the $J_L$ dependence of $\varepsilon_T(J_L,J_F)$ for the indicated $J_F$ at $2N = 16$ and $24$ in the upper and lower panels, respectively. The ladder with $-J_F = 1$ in Eq. \ref{ssm_eq1} is a $J_1-J_2$ model with $J_1 = 1$ and $J_2 = -J_L$. Ladders with other $J_F$ have two spins per unit cell and correspond to alternating $J_1-J_2$ models with $J_1 = (1 - J_F)/2$ and alternation $\pm (1 + J_F)/2$. Since the $J_1-J_2$ model has noninteracting legs at $J_1 = 0$, the gap at $J_F = 1$ is entirely due to alternation. The gaps at $J_F = 0$ and $1/2$ in Fig. \ref{fig5_p} are equal at $1/J_L = 0$. As expected, alternating exchanges increase the gap when $J_L$ is finite.

Eq. \ref{ssm_eq1} conserves total $S$ but not the spins $S_A = S_B \leq N/2$ of each leg. Equal $J^\prime$ between all spins in different legs leads to separately conserved $S$, $S_A$ and $S_B$. Angular momentum addition returns $\varepsilon_T = J^\prime$ when $J^\prime > 0$. The mean-field approximation for  $-J_F$ and $J_A = 1$ is $J^\prime = (1 - J_F)/N$. The same result holds for the $J_1-J_2$ model with $2N$ exchanges $J_1 = (1 - J_F)/2$. In the limit $J_L \to \infty$, the gap at system size $2N$ is
\begin{equation}
{\varepsilon_T(J_F,2N)}= (1-J_F)/N,    \hspace{0.8cm} J_F\leq 1,J_L\to \infty.
\label{ssm_eq9}
\end{equation}   
Eq. \ref{ssm_eq9} agrees quantitatively with the numerical results at $1/J_L = 0$ in Fig. \ref{fig5_p} for ladders and $J_1-J_2$ models. Alternation increases $\varepsilon_T$.

The mean-field approximation has apparently not been recognized in systems with F exchange $-J_L$ in legs. The ladder with $J_F = 0$ and $J_A > 0$ in Fig. \ref{fig1_p} has been studied numerically \cite{Roji1996} and field theoretically \cite{Kolezhuk1996}. The F state is unconditionally unstable when $J_A > 0$, as expected on general grounds; it can be stabilized \cite{Roji1996,Kolezhuk1996} by an Ising contribution to the isotropic exchange $-J_L$. Eq. 9 is consistent with general expectations and provides quantitative gaps for finite ladders with $-J_F \leq 1$ in Eq. \ref{ssm_eq1}. The mean-field $\varepsilon_T(J_F,2N)$ is elementary at $1/J_L = 0$ and rigorously decreases as $1/N$. It is a good approximation to at least $1/J_L = 0.1$.

We conclude this Section by highlighting the difference between no net AF exchange and no exchange. The ladder at $J_F \to \infty$ is a spin-1 HAF with $J = (1 - 2J_L)/4$. The singlet/F boundary is at $J = 0$ where the energy per dimer is $\varepsilon_0 = \varepsilon_F = -J_F/4$. Since the boundary in Eq. \ref{ssm_eq5} is at $2J_L > 1$ when $J_F$ is finite, the ground state is a singlet at $1 - 2J_L = 0$ and no net exchange, with per dimer energy
\begin{equation}
{\varepsilon_0(1/2,J_F)/J_F}=-1/4-c/J_F^2.
\label{ssm_eq10}
\end{equation} 
The first-order energy of Eq. \ref{ssm_eq8} is zero at $2J_L = 1$. There is a second-order correction because $J_A = 1$ and $-J_L = 1/2$ are between different spins, $2r$, $2r + 1$ for $J_A$ and $r$, $r + 2$ for $J_L$. Eq. \ref{ssm_eq10} holds for $J_F > 10$ with $c = 0.516$ at both $2N = 16$ and $24$.

The $J_L \to \infty$ limit of the ladder is a $J_1-J_2$ model. The singlet/F boundary is at $J_1 = 0$ with $\varepsilon_0 = -J_L/2$ when $J_F = 1$. The boundary of the ladder, $2J_L = J_F/(J_F - 1)$, is at $J_F > 1$ when $J_L$ is finite. The ground state is a singlet at $J_F = 1$ and no net exchange between legs. But $\varepsilon_0(J_L,1)$ has second-order corrections in $1/J_L$ since $J_A$ and $J_F$ are between different spins. We find $\varepsilon_0(J_L,1)/J_L = -1/2  - d/J_L^2$ for $J_L > 10$ with $d = 0.271$ and $0.282$ at $2N = 16$ and $24$.

\section{String correlation functions \label{sec-IV}}

We obtain an explicit relation between spin-1 and spin-1/2 string correlation functions. Girvin and Arovas define \cite{Girvin1989} string correlation functions of consecutives $s = 1$ spins as
\begin{equation}
{\tilde{g}}(n)=-\langle s_1^z (expi\pi \sum_{j=2}^{n} {s}_j^{z}) {s}_{n+1}^{z} \rangle.
\label{ssm_eq11}
\end{equation}
The expectation value is with respect to the singlet ground state in the thermodynamic limit or in finite chains with periodic boundary conditions. The $(n - 1)$ spins-1 in the exponent can be written as $2(n - 1)$ spins-1/2 with $s_j = S_{2j-1} + S_{2j}$. The other spins are $s_1 = S_1 + S_2$ and $s_{n+1} = S_{2n +1} + S_{2n + 2}$. To have strings of consecutive spins-1/2, Hida \cite{Hida1992} chose spins $S_2$ and $S_{2n+1}$ and used the spin-1/2 identity,
\begin{equation}
{-4S_m^zS_n^z}=expi\pi(S_m^z+S_n^z)
\label{ssm_eq12}
\end{equation}
to shift those spins into the exponent. This leads to $g_2(N)$ with $N$ consecutive spins from $2$ to $N + 1$ in Eq. \ref{ssm_eq2}. As anticipated \cite{Hida1992} on including the factor of $4$, the string order $g_2(\infty)$ in the limit $J_F \to \infty$ is equal to $\tilde{g}(\infty)$. 

The spin-1/2 string correlation function defined in Eq. \ref{ssm_eq2} is not limited to either $J_F \to \infty$ or $N \to \infty$. Systems with two spin spins per unit cell have two strings of $N$ spins. The string $g_1(N)$ in Eq. \ref{ssm_eq2} has $N/2$ exchanges $-J_F$ and $N/2-1$ exchanges $J_A = 1$ while $g_2(N)$ has $N/2$ exchanges $J_A$ and $N/2-1$ exchanges $-J_F$. The size dependencies of $g_1(N)$ and $g_2(N)$ are very different.

The string operator $\hat{g}_1(N)$ has an even number $N$ of consecutive spins from $1$ to $N$. Singlet VB diagrams $\lvert q \rangle$ in systems of $2N$ spins have $N$ lines $(m,n)$ that correspond to singlet-paired spins in Eq. \ref{ssm_eq3}. Repeated use of Eq. \ref{ssm_eq12} leads to
\begin{eqnarray}
\label{ssm_eq13}
exp (i\pi \sum_{j=1}^{N} {S}_j^{z}) \lvert q \rangle=\lvert q \rangle, \hspace{0.4cm} 1\leq m, n\leq N  \nonumber \\ 
& & \hspace*{-110pt} = \lvert q \rangle_T,  \hspace{0.4cm} otherwise.
\end{eqnarray}
Diagrams $\lvert q \rangle$ with $1 \leq m,n \leq N$ are eigenfunctions of  with unit eigenvalue. The factor of $4$ in Eq. \ref{ssm_eq12} is required for normalization, $\langle q\rvert q\rangle = 1$. The eigenfunctions are all possible singlets based on spins in the string. 

Diagrams $\lvert q\rangle$ that are not eigenfunctions contain one or more pairs of bridging lines $(m,n)$ with only one spin in the string. Then $\hat {g}_1(N) \lvert q\rangle$ generates a diagram $\lvert q\rangle_T$ with triplet-paired spins $(m,n)_T = (\alpha_m \beta_n + \beta_m \alpha_n)/\sqrt{2}$ at all bridging lines. Spin orthogonality ensures $\langle q\rvert q\rangle_T = 0$ but finite $\langle q^ \prime \rvert q\rangle_T$ is possible with other singlets $\lvert q^ \prime \rangle$. The Appendix summarizes two general properties of singlet VB diagrams, overlaps and dimensions. Overlaps $S_{q^\prime q} = \langle q^ \prime \rvert q\rangle$ are needed to evaluate expectation values. $R_0(2N)$ in Table \ref{tab:table2} is the number of singlet diagrams at system size $2N$. A string of $N$ spins has $R_0(N)$ eigenfunctions, each $R_0(N)$-fold degenerate, without any bridging lines. The relative number of diagrams with bridging lines increases with system size as indicated by the decreasing ratio $R_0(N)^2/ R_0(2N)$ in Table \ref{tab:table2}.  

\begin{table}
\normalsize
\caption{\label{tab:table2} $R_0(2N)$ is the number of singlet VB diagrams for $2N$ spins. $R_0(N)$ is the number of eigenstates of an N-spin string. Eq. \ref{ssm_eq20} is Stirling's approximation.}
\begin{center}
\begin{tabular}{|C{1.0cm}|C{1.4cm}|C{1.4cm}|C{2.5cm}|C{1.4cm}|}
\hline
$2N$ & ${R_0(2N)}$ & ${R_0(N)}$ & ${R_0(N)^2/R_0(2N)}$ & Eq. \ref{ssm_eq20} \\ 
\hline
$12$ & $132$ & $5$ & $0.189$ &  $0.199$ \\
\hline
$16$ & $1430$ & $14$ & $0.137$ &  $0.143$ \\
\hline
$20$ & $16796$ & $42$ & $0.105$ &  $0.109$ \\
\hline
$24$ & $208012$ & $132$ & $0.0838$ &  $0.0861$ \\
\hline
\end{tabular}
\end{center}
\end {table}

\begin{figure}[h!]
\includegraphics[width=\linewidth]{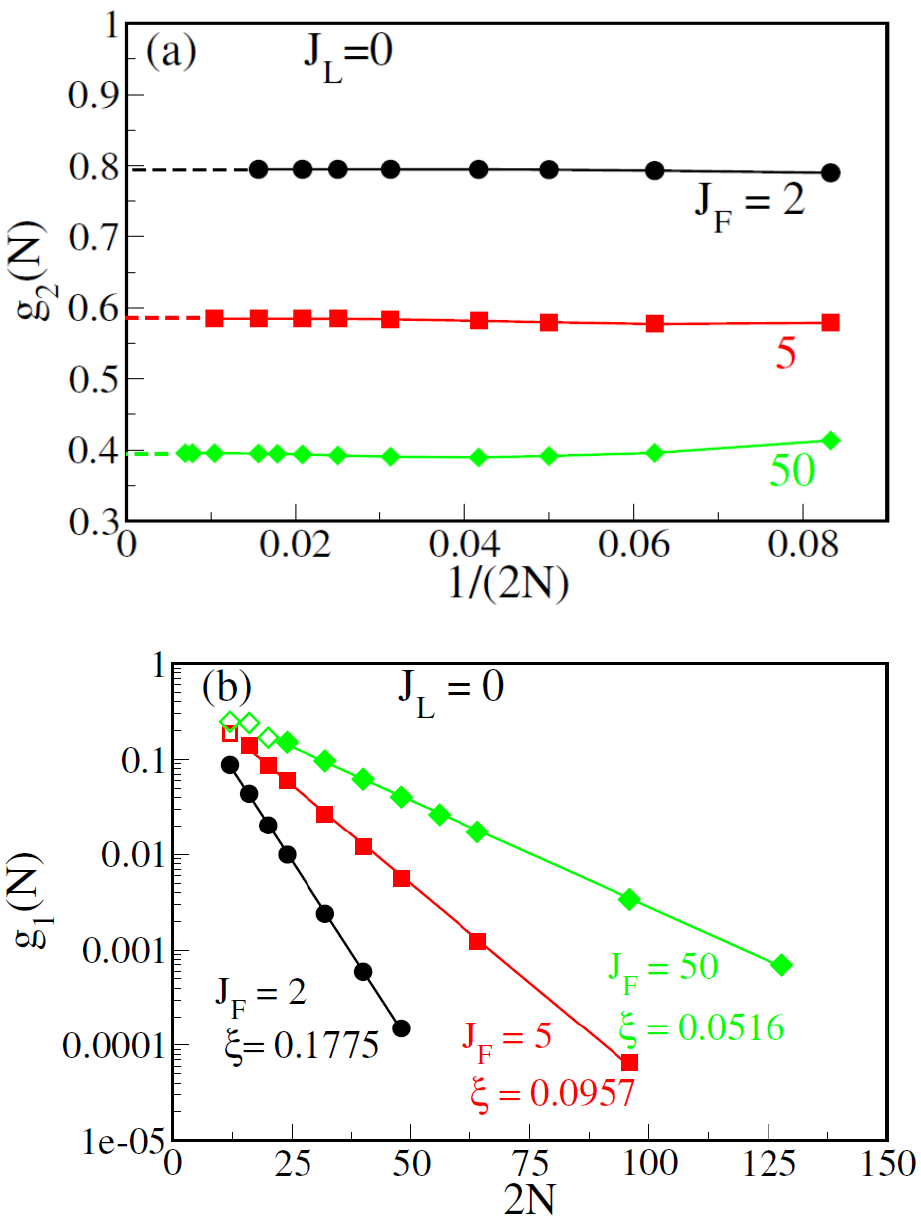}
\caption{(a) String correlation functions $g_2(N)$ at system size $2N$, $J_L = 0$ and $J_F = 2, 5$ and $50$; linear extrapolation to string order $g_2(\infty)$. (b) $g_1(N)$ for the same $J_L$, $J_F$ with solid symbols for $N \geq N^*$, the minimum of $g_2(N)$, open symbols for $N < N^*$. The lines are Eq. \ref{ssm_eq15} with the indicated $\xi$.}
\label{fig6_p}
\end{figure}

We compute the string correlation functions $g_2(N)$ and $g_1(N)$ of the F-AF ladder with $2N$ spins  Eq. \ref{ssm_eq1}. The upper panel of Fig. \ref{fig6_p} shows the size dependence of $g_2(N)$ as $1/(2N)$ from $2N = 12$ to $144$ at $J_L = 0$ and $J_F = 2$, $5$ and $50$. The lower panel shows $g_1(N)$ on a logarithmic scale. At any system size, $g_2(J_F)$ is larger than $g_1(J_F)$ and decreases with $J_F$ while $g_1(J_F)$ increases with $J_F$. Although not evident on the scale of Fig. \ref{fig6_p}, $g_2(N)$ at $J_F = 2$ increases from $0.789978$ at $2N = 12$ and extrapolates to $g_2(\infty) = 0.794918$. At $J_F = 5$ and $50$, $g_2(N)$ has a shallow minimum at $N^* = 8$ and $12$, respectively. Convergence to $g_2(\infty)$ is again from below.

The following statements summarize results for other parameters $J_L$, $J_F$. String correlation functions satisfy the inequality,
\begin{equation}
1\geq g_2(N) > g_1(N) \geq 0. 
\label{ssm_eq14}
\end{equation}
The function $g_2(N)$ has a shallow minimum at system size $2N^*$. The size dependence of $g_1(N)$ is exponential for $N \geq N^*$
\begin{equation}
{g_1(N)}=g_1(N^*)exp[-2{\xi} (N-N^*)], \hspace{0.4cm} N\geq N^*.
\label{ssm_eq15}
\end{equation} 
The $g_2(N)$ minima in Fig. \ref{fig6_p} are $N^* = 4$, $8$ and $12$, respectively, for $J_F = 2$, $5$ and $50$. The lines for $N \geq N^*$ in the lower panel are Eq. \ref{ssm_eq15} with the indicated $\xi$.

We interpret $g_2(N)$ and $g_1(N)$ in terms of the VB ground state, Eq. \ref{ssm_eq4}, with coefficient $C(q)$ for diagram $\lvert q \rangle$. The exact ground state is $\lvert K2 \rangle$ when $J_L = J_F/2 \leq 1$, with $C(K2) = 1$ and $C(q) = 0$ for all other $\lvert q \rangle$. The expectation values are $g_2(N) = 1$ and $g_1(N) = 0$ independent of system size. The shortest string is $N = 4$ since consecutive spins return the spin correlation function $-4\langle {S_1^z} {S_2^z}\rangle$.

\begin{figure}[h!]
\includegraphics[width=\linewidth]{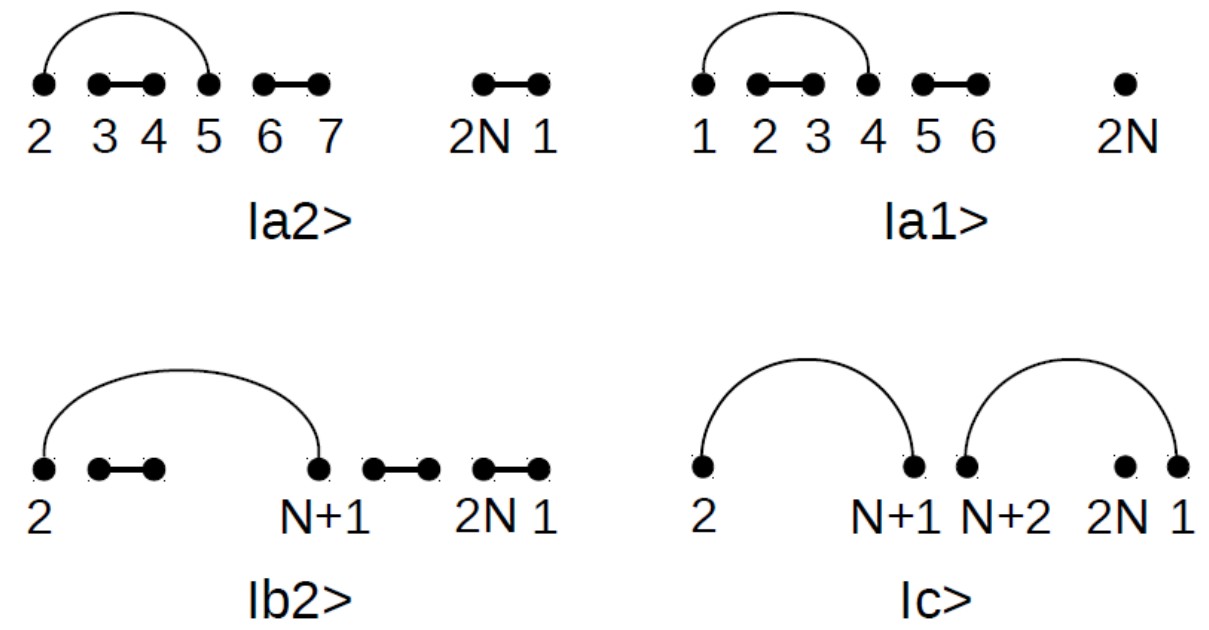}
\caption{Representative singlet VB diagrams $\lvert q \rangle$ with $N$ lines $(m,n)$ in ladders with $2N$ spins and $C_N$ translational symmetry: $\lvert a2 \rangle$ has one line $(2,5)$ of length $3$ and $(N-1)$ lines of unit length; $\lvert a1\rangle$ has one line $(1,4)$ of length $3$; $\lvert b2 \rangle$ has one line $(2,N + 1)$ of length $N - 1$, the maximum length; $\lvert c\rangle$ has two lines of length $N - 1$, $(2,N + 1)$ and $(N + 2,1)$.}
\label{fig7_p}
\end{figure}

The ground state is a linear combination of singlet diagrams $\lvert q \rangle$ when $\lvert J_L - J_F/2 \rvert> 0$. The representative singlet VB diagrams $\lvert q \rangle$ in Fig. \ref{fig7_p} have $N$ lines $(m,n)$; lines not shown explicitly are between neighbors $(m,m + 1)$. The diagram $\lvert a2 \rangle$  differs from $\lvert K2 \rangle$ by two lines, $(2,5)$ and $(3,4)$. The $N$ symmetry-related diagrams with a line $(2r,2r + 3)$ have equal $C(q)$. The $N$ diagrams $\lvert a1 \rangle$ in Fig. \ref{fig7_p} with a line $(2r - 3,2r)$ have equal $C(a1) < C(a2)$ since only line $(2,3)$ is shared with $\lvert K2 \rangle$. Diagram $\lvert b2 \rangle$ has a line $(2r,2r + N - 1)$ of length $N - 1$, the longest at system size $2N$. Diagram $\lvert c \rangle$ has two lines of length $N - 1$ and $(N - 2)$ lines of unit length.

Table \ref{tab:table3} shows the size dependence of selected coefficients $C(q)$ at $J_L = 0$, $J_F = 5$. The strong decrease of $C(K1)$ with system size is due to the overlap $\langle K1\rvert K2\rangle =(-2)^{-(N + 1)}$. The decrease of $C(a1)$ is also due to overlap. As shown in the Appendix, diagrams $\lvert q \rangle$ that differ from $\lvert K2 \rangle $ by a finite number of lines are asymptoticly orthogonal to diagrams $\lvert q^\prime \rangle$ that differ from $\lvert K1 \rangle$ by a finite number of lines. The size dependence of $C(b2)$ illustrates the range of spin correlations, which is short at $J_L = 0$, $J_F = 5$, consistent with large 
$\varepsilon_T (0,5)$ in Fig. \ref{fig3_p}. Diagram $\lvert c\rangle$ has two lines of maximum length, and short-range spin correlations explain its size dependence.

\begin{table}
\normalsize
\caption{\label{tab:table3}Ground-state coefficient $C(q)$ of diagrams $\lvert q\rangle$ in Eq. 1 with $J_L = 0$, $J_F = 5$ and $2N$ spins. $\lvert K1\rangle$ and $\lvert K2\rangle$ are shown in Fig. \ref{fig1_p}, and $\lvert a2\rangle$, $\lvert a1\rangle$, $\lvert b2\rangle$ and $\lvert c\rangle$ in Fig. \ref{fig7_p}.}
\begin{center}
\begin{tabular}{|C{1.9cm}|C{1.9cm}|C{1.9cm}|C{1.9cm}|}
\hline
$C(q) \backslash 2N$ & $8$ & ${12}$ & ${16}$ \\ 
\hline
$C(K2)$ & $0.7583$ & $0.6454$ & $0.5607$ \\
\hline
$C(K1)$ & $0.1188$ & $0.0217$ & $0.0034$ \\
\hline
$C(a2)$ & $0.2981$ & $0.2545$ & $0.2213$ \\
\hline
$C(a1)$ & $0.0763$ & $0.00556$ & $0.00027$ \\
\hline
$C(b2)$ & $0.2981$ & $0.0833$ & $0.00045$ \\
\hline
$C(c)$ & $0.1172$ & $0.0109$ & $0.00144$ \\
\hline
\end{tabular}
\end{center}
\end{table}

Turning to string correlation functions, we note that $(N - 2)$ of the diagrams $\lvert a2 \rangle$ are eigenfunctions of $\hat{g}_2(N)$ while two diagrams have bridging lines, either $(2N,3)(1,2)$ or $(N,N + 3)(N + 1,N + 2)$. The relative number of bridging lines in $\lvert a2\rangle$ decreases with system size. On the other hand, only two diagrams $\lvert b2\rangle$ are eigenfunctions of $\hat{g}_2 (N)$; the other $(N - 2)$ have bridging lines. As seen in Table \ref{tab:table2}, the relative number of diagrams with bridging lines increases with system size, and so do their coefficients $C(q)$ for parameters $J_L$, $J_F$ that increase the range of spin correlations. 

Since the ladder is gapped, spin correlations are finite ranged and $C(q)$ must be small for diagrams with lines $(m,n)$ that exceed the range. We suppose $N^*$ to be an estimate of the range. Then $g_2(N)$ and $g_1(N)$ decrease with system size up to $2N^*$ because the bridging lines increase more rapidly than eigenstates in Table \ref{tab:table2}. By hypothesis, $C(q)$ is negligible for diagrams with line longer than $N^*$. Then $g_2(N)$ increases when $N > N^*$ because diagrams with lines shorter than $N^*$ are only bridging at the ends of increasingly long strings. The range $N^*$ limits finite $C(q)$ to diagrams that differ from $\lvert K2\rangle$ by a specified number of lines. The exponential decrease of $g_1(N)$ for $N > N^*$ in Eq. \ref{ssm_eq15} is consistent with the asymptotic orthogonality of diagrams such as $\lvert a2\rangle$ and $\lvert a1\rangle$ that differ from $\lvert K2\rangle$ and $\lvert K1\rangle$, respectively, by two lines. It follows that $g_1(\infty)=0$ and that $C(q) = 0$ in the thermodynamic limit for the $R_0(2N)/2$ diagrams $\lvert q\rangle$ whose squared overlap is larger with $\lvert K1\rangle$ than with  $\lvert K2\rangle$.

\begin{figure}[h!]
\includegraphics[width=\linewidth]{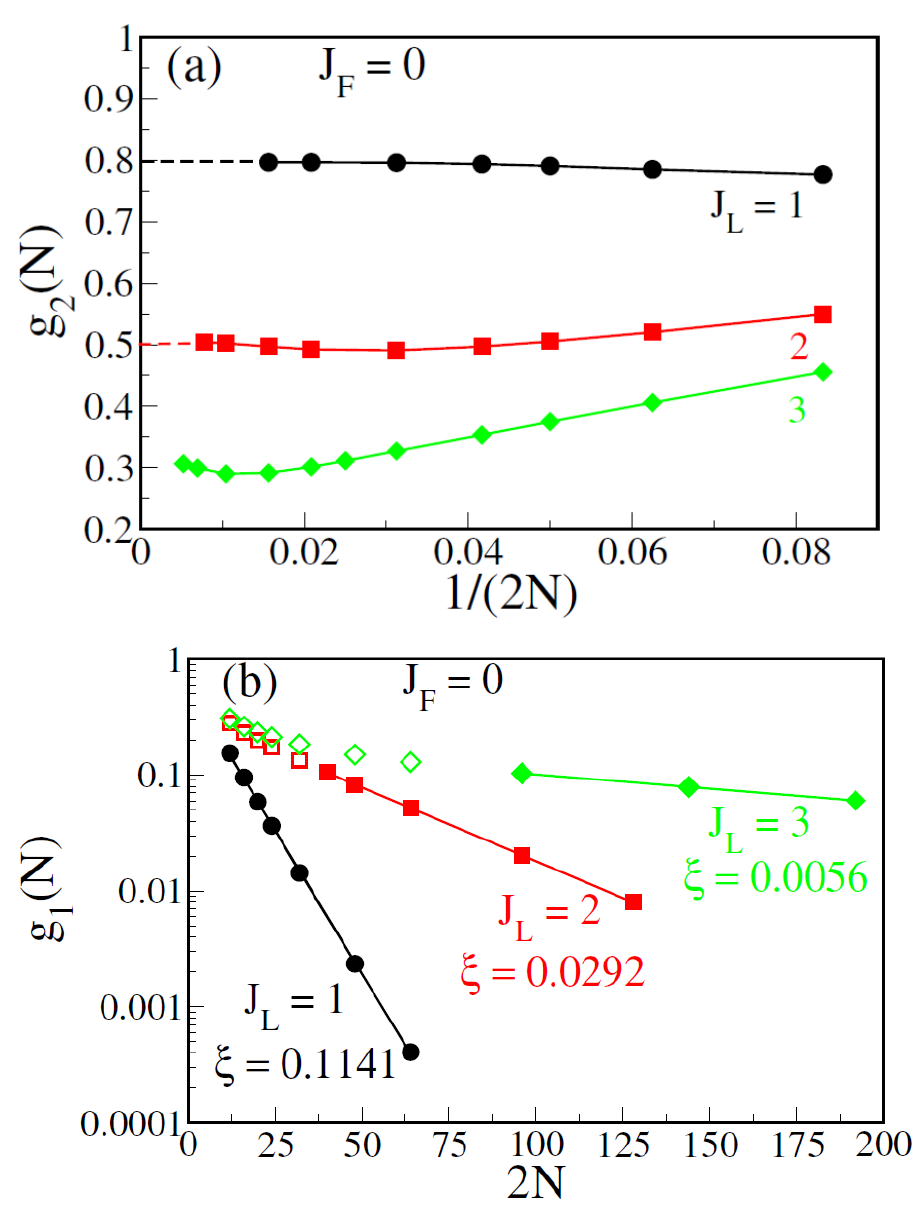}
\caption{(a) String correlation functions $g_2(N)$ at system size $2N$ and $J_F = 0, J_L = 1, 2$ and $3$. Linear extrapolation to string order $g_2(\infty)$. (b) $g_1(N)$ for the same $J_L$, $J_F$; solid symbols for $N \geq N^*$, open symbols for $N < N^*$. The lines are Eq. \ref{ssm_eq15} with the indicated $\xi$.}
\label{fig8_p}
\end{figure}

The panels of Fig. \ref{fig8_p} show the size dependence of $g_2(N)$ and $g_1(N)$ in ladders with $2N$ spins, $J_F = 0$ and $J_L = 1$, $2$ and $3$. The $g_2(N)$ minima are $N^* = 4$, $16$ and $48$ for $J_L = 1$, $2$ and $3$. Larger systems are required for accurate extrapolation of $g_2(N)$ at $J_L = 3$. The solid lines $g_1(N)$ in the lower panel are Eq. 15 with the indicated $\xi$. 


Table \ref{tab:table4} lists the string order $g_2(\infty)$, the minimum $N^*$ of $g_2(N)$ and the spin gap $\varepsilon_T$ for representative parameters $J_L$, $J_F$. We recall that $g_2(\infty) = 1 = C(K2)$ when $J_L = J_F/2 \leq 1$ while $\varepsilon_T$ decreases from $1$ at the origin to $\approx 0.003$ at the singlet/F boundary. A $g_2(N)$ minimum at $N^*$ requires $J_L$, $J_F$ that lead to significant $C(q)$ for diagrams $\lvert q \rangle$ with lines $(m,n)$ longer than $4$, the shortest string. The first two entries at $\lvert J_L - J_F/2 \rvert = 0.25$ have almost equal $g_2(\infty)$ but quite different gaps; $g_2(N)$ has a minimum at $2N = 8$ for the smaller $\varepsilon_T$ but not for the larger one, and the coefficients $C(K2)$, $C(a2)$ are by far the largest in either case. Systems with $N^* = 4$ or $8$ in the Table return $g_2(\infty) > 0.75$ or $> 0.5$, respectively. The exponential decrease of $g_1(N)$ for $N \geq N^*$ in Figs. \ref{fig6_p} (b) and \ref{fig8_p} (b) starts around $g_1(N^*) \approx 0.1$. The interpretation is that $C(q)$ is small for diagrams with lines $(m, n)$ longer than $N^*$. String order $g_2(\infty) < 0.25$ indicates longer-ranged spin correlations with $N^* > 50$ and gaps $\varepsilon_T < 0.1$.

\begin{table}
\normalsize
\caption{\label{tab:table4}String order $g_2(\infty)$, minimum $N^*$ of $g_2(N)$ and gap $\varepsilon_T$ in the thermodynamic limit for $J_L$, $J_F$ in Eq. \ref{ssm_eq1}.}
\begin{center}
\begin{tabular}{|C{1.0cm}|C{1.4cm}|C{1.4cm}|C{2.5cm}|C{1.4cm}|}
\hline
$J_L$ & $J_F$ & $g_2(\infty)$ & ${N^*}$  & $\varepsilon_T$ \\ 
\hline
\hline
$0.25$ & $1$ & $0.98144$  &  $-$ & $0.777$   \\
\hline
$1$ & $1.5$ & $0.98143$  & $4$ & $0.242$   \\
\hline
$1$ & $1$ & $0.941$  & $4$  & $0.340$ \\
\hline
$0.5$ & $0$ & $0.926$ &$4$ & $0.610$ \\
\hline
$1$ & $0$ & $0.797$ &$4$ & $0.370$ \\
\hline
$0$ & $2$ & $0.795$ &$4$ & $0.416$ \\
\hline
$0.5$ & $5$ & $0.675$ &$6$ & $0.183$ \\
\hline
$0.25$ & $5$ & $0.634$ &$6$ & $0.196$ \\
\hline
$0$ & $5$ & $0.585$ &$8$ & $0.225$ \\
\hline
$2$ & $0$ & $0.503$ &$16$ & $0.110$ \\
\hline
$2$ & $0.5$ & $0.490$ &$24$ & $0.053$ \\
\hline
$2$ & $1$ & $<0.09$ & $>100$ & $0.016$ \\
\hline
$0$ & $50$ & $0.396$ & $12$ & $0.11$ \\
\hline
$3$ & $0$ & $ \sim {0.31}$  & $48$ & $0.038$ \\
\hline
\end{tabular}
\end{center}
\end{table}

The $g_2(N^*)$ minimum is exceptionally shallow at $J_L = 0.5$ and $J_F = 5$. Short-range correlations are to be expected at zero net exchange $1 - 2J_L$ between F rungs. We find constant $g_2(N)$ for $N \geq N^*$ to three decimal places, very small $g_1(N^*) < 0.02$ and the only deviation from exponential behavior seen so far. Spin correlations at $J_F = 5$ in Table \ref{tab:table4} are shorter-ranged at $J_L = 0.5$ than at $J_L = 0$. 

There is no net exchange between legs when $J_F = 1$. The $J_L = 2$ gaps in Table \ref{tab:table4} decrease from $J_F = 0$ to $J_F = 1$. In contract to zero net exchange between rungs, however, $N^*$ at $J_L = 2$ increases from $16$ at $J_F = 0$ to $24$ at $J_F = 0.5$ and exceeds $100$ at $J_F = 1$.
Longer ladders than $2N \approx 200$ will be required for $g_2(\infty) < 0.25$. For example, $g_2(N)$ is still decreasing at $2N = 192$ at $J_L = 1.5$, $J_F = 1.3$ or at $J_F = 1.5$, $J_L = 1.35$, the parameters with $\varepsilon_T < 0.01$ in Fig. \ref{fig4_p}.
\begin{figure}[h!]
\includegraphics[width=\linewidth]{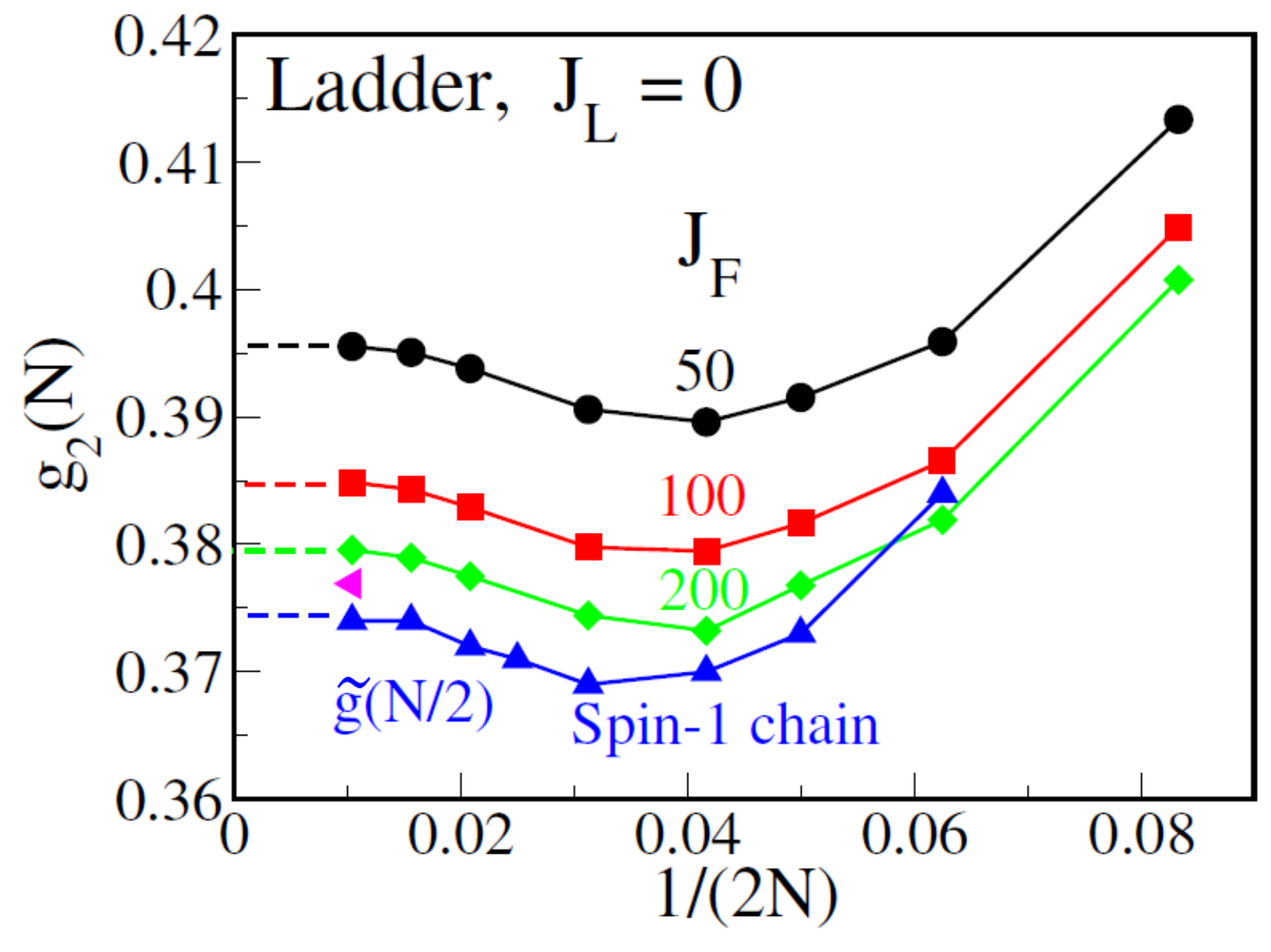}
\caption{ Size dependence of $g_2(N)$ at $J_L = 0$ and the indicated $J_F$ from $2N = 12$ to $96$. The string correlation function $\tilde{g}(N/2)$ is for the HAF with $N$ spins-1. The magenta point at $2N = 96$ is for $J_F = 400$.
}
\label{fig9_p}
\end{figure}

We now turn to the $J_F \to \infty$ limit of the ladder, the spin-1 HAF with $J = 1/4$ at $J_L = 0$. Fig. \ref{fig9_p} compares the size dependence of $g_2(N)$ at system size $2N$, $J_L = 0$, and $J_F$ with $\tilde{g}(N/2)$, Eq. \ref{ssm_eq11}, the string correlation function for $N$ spins-1. Note the expanded scale. The string orders $\tilde{g}(\infty)$ and $g_2(\infty)$ are equal in the limit $J_F \to \infty$ by construction \cite{Hida1992}, as discussed above, but $g_2(N)$ is not equal to $\tilde{g}(N/2)$ at either finite $J_F$ or finite $N$. The $\tilde{g}(N/2)$ minimum occurs at $N = 16$ spins-1. The string order is \cite{Huse1993} $\tilde{g}(\infty) = 0.374325$ while we obtain $\tilde{g}(\infty)= 0.37427$ at $48$ spins-1. We find $g_2(48) = 0.37692$ for $96$ spins-1/2 and $J_F = 400$, and string order $0.37427$ on linear extrapolation to $1/J_F = 0$.


The $\tilde{g}(N/2)$ minimum is a new result. Previous studies considered $\tilde{g}(p)$ at constant system size $n$ and hence constant spin correlations with periodic \cite{Girvin1989} or open \cite{Huse1993} boundary conditions; $\tilde{g}(p,n)$ decreases with $p$ up to $n/2$ in small cyclic systems or to $\tilde{g}(\infty)$ as shown in Fig. \ref{fig5_p} of ref. \cite{Huse1993}. The corresponding spin-1/2 function $g_2(2p,2N)$ has variable $p$ at constant system size $2N$, and it also decreases to $2p = N$ or to $g_2(\infty)$ in the thermodynamic limit. We have instead studied the size dependence of $g_2(N,2N)$ and found an unanticipated minimum at $N^*$. Convergence to string order $g_2(\infty)$ is from below. The VB analysis rationalizes the size dependencies of both $g_2(N,2N)$ and $\tilde{g}(n/2,n)$.


Finite string order $g_2(\infty)$ and gap $\varepsilon_T$ are expected on general grounds in dimerized ladders with $–J_F \ne J_A$ and two spins per unit cell. The ground state is a BOW. The limit $J_F \to \infty$ generates inversions centers at the centers of rungs, and the ladder becomes a spin-$1$ HAF with $J = (1 - 2J_L)/4 > 0$ and $Z_2$ symmetry.


\section{Spontaneous dimerization \label{sec-V}}

The F-AF ladder, Eq. \ref{ssm_eq1}, has two string correlation functions with an even number $N$ of consecutive spins-$1/2$. The nondegenerate ground state is a BOW due to alternate first-neighbor exchanges $-J_F$ and $J_A = 1$. We set $-J_F = 1$ in this Section and discuss the $J_1-J_2$ model with $J_1 = 1$ and $J_2 = -J_L$. The model has one spin per unit cell, $C_{2N}$ translational symmetry, and inversion symmetry $\sigma$ at sites. The ground state for $2N$ spins, $N$ even is odd under inversion, $\sigma = -1$. The coefficients of $\lvert K1\rangle$ and $\lvert K2\rangle$ or of $\lvert a2\rangle$ and $\lvert a1 \rangle$ in Table \ref{tab:table3} are then equal with opposite sign. We have $C(q^\prime) \pm C(q)$ for symmetry-adapted linear combinations of singlets $\lvert q \rangle$ and $\lvert q^\prime \rangle =\sigma\lvert q \rangle$.  The lowest singlet excited state has $\sigma = 1$ symmetry and $C(q^\prime) = C(q)$. 

\begin{figure}[h!]
\includegraphics[width=\linewidth]{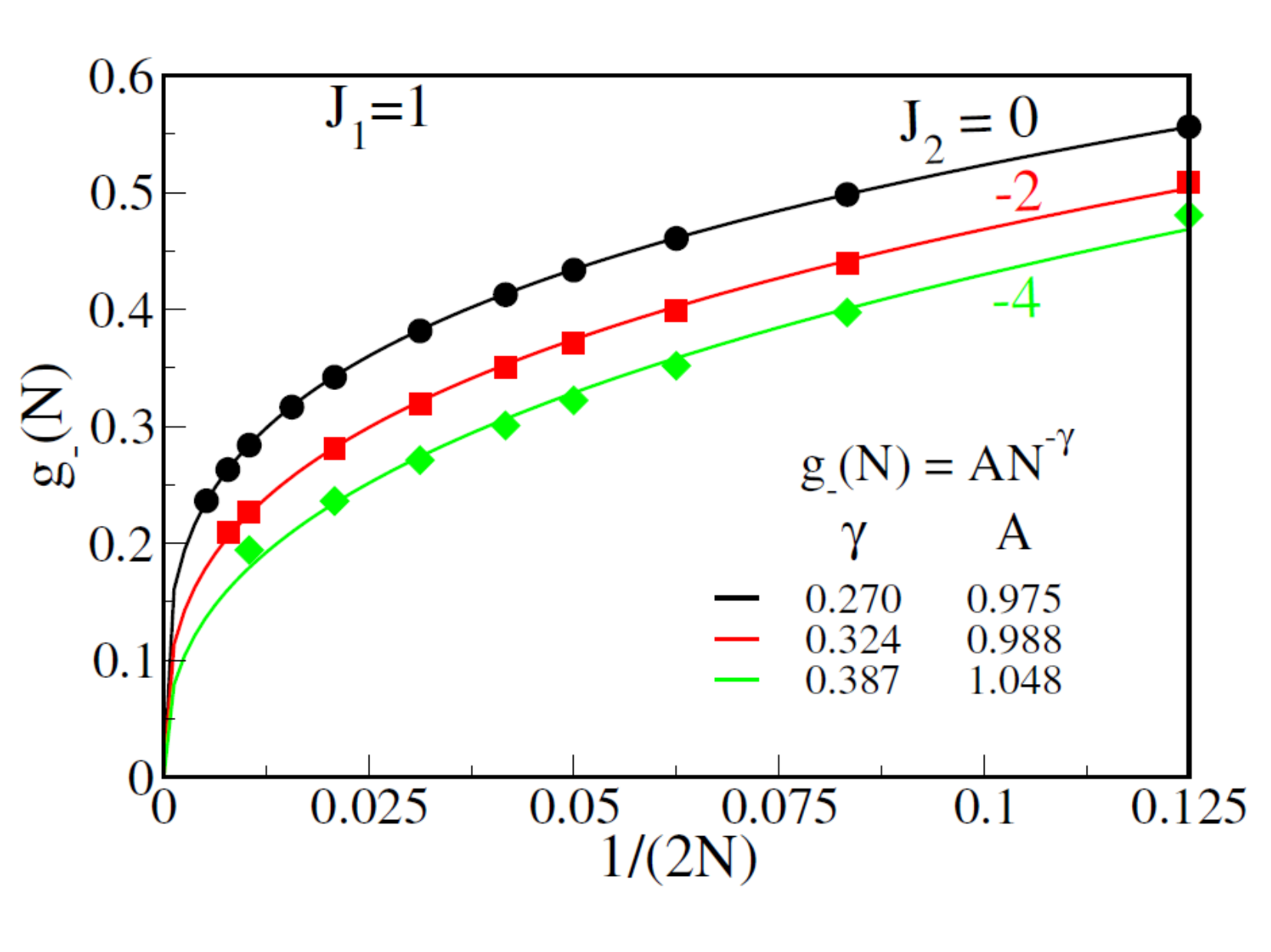}
\caption{Size dependence of the string correlation function $g_-(N)$ of $J_1-J_2$ models with $2N$ spins, $J_1 = 1$ and $J_2 = 0, -2$ and $-4$. The $J_2 = 0$ expression \cite{Hida1992} is used at $J_2 = -2$ and $-4$.
}
\label{fig10_p}
\end{figure}

The string correlation function $g_-(N)$ is the ground-state expectation value. Hida \cite{Hida1992} applied field theory to the spin-$1/2$ HAF $(J_2 = 0)$ and concluded that $g_-(N)$ is proportional to $N^{-1/4}$, consistent with ED up to $2N = 24$. Fig. \ref{fig10_p} shows the size dependence of $g_-(N)$ at $J_2 = 0$ from $2N = 12$ to $192$, The exponent $\gamma(0) = 0.270$ is the best fit, in good agreement with field theory. Since the model with $J_2 < 0$ is not frustrated, the size dependence of $g_-(N)$ at $J_2 = -2$ and $-4$ in Fig. \ref{fig10_p} is also fit as $AN^{-\gamma}$.

The $J_1-J_2$ model is frustrated when $J_2 > 0$. The ground state is doubly degenerate in the dimer phase \cite{Soos2016} $J_c = 0.2411 \leq J_2 \leq 1/2$. In finite systems, the singlets $\sigma = -1$ and $+1$ are the ground and first excited states, respectively. They are degenerate at $J_2 = 1/2$, the Majumdar-Ghosh point \cite{Majumder1969}, where the exact $\sigma = \pm 1$ ground states are the plus and minus linear combinations of $\lvert K1 \rangle$ and $\lvert K2 \rangle$. The system is spontaneously dimerized.

The broken-symmetry state $\lvert K2 \rangle$ at $J_2 = 1/2$ returns $g_2(N) = 1$, $g_1(N) = 0$ as discussed for the ladder while $\lvert K1\rangle$ has $g_1(N) = 1$, $g_2(N) = 0$. Due to overlaps, the string correlations functions $g_\pm(N)$ are size dependent. A straightforward calculation leads to
\begin{equation}
{g_-(N)}=1/2+1/(2^N+2).
\label{ssm_eq16}
\end{equation} 
The $g_+(N)$ expression has minus signs in Eq. \ref{ssm_eq16}. Convergence to the thermodynamic limit is exponential.

\begin{figure}[h!]
\includegraphics[width=\linewidth]{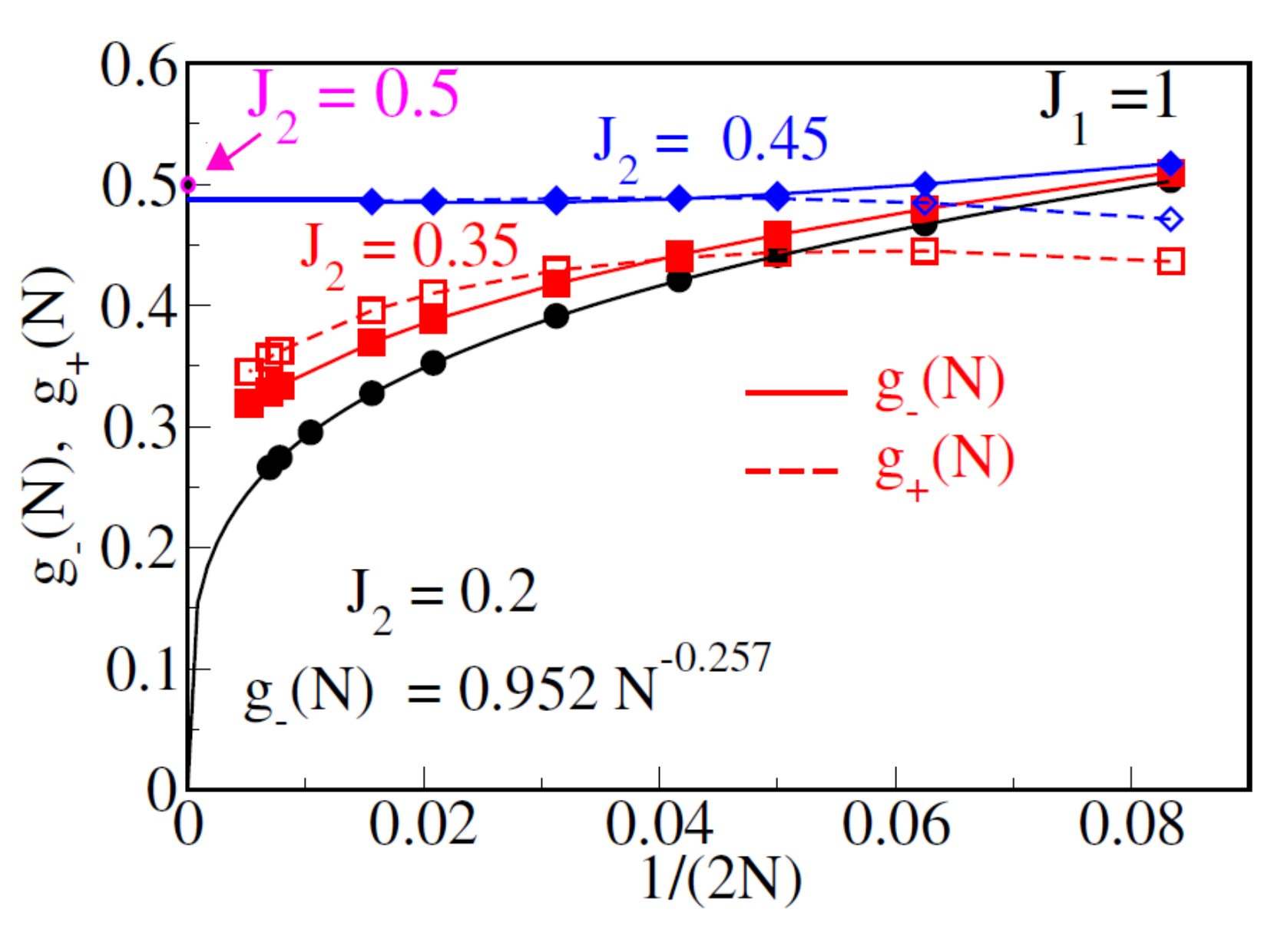}
\caption{String correlation function $g_-(N)$ at $J_2 = 0.2$ in the gapless phase, fit as in Fig. \ref{fig10_p}. Solid and dashed lines are $g_-(N)$ and $g_+(N)$ at $J_2 = 0.35$ and $0.45$ in the gapped dimer phase. The point at $J_2 = 0.5$ is exact. The lines at $J_2 = 0.35$ are to guide the eye.}

\label{fig11_P}
\end{figure}

Fig. \ref{fig11_P} shows the size dependence of $g_-(N)$ at $J_2 = 0.20 < J_c$ in the gapless phase and both string correlation functions at $J_2 = 0.45$ and $0.35$ in the dimer phase.  
The $g_-(N)$ points at $J_2 = 0.20$ are fit to $AN^{-\gamma}$ as in Fig. \ref{fig10_p} with $\gamma = 0.257$. The point at $0.50$ is exact for $J_2 = 0.5$. The $g_-(0.45,N)$ and $g_+(0.45,N)$ curves cross twice before converging to string order $g_-(\infty) = g_+(\infty) = 0.485$. Rapid size convergence and slightly reduced string order are expected close to $J_2 = 1/2$. We did not anticipate curve crossing; the ground state is odd under inversion at all system sizes.

The $J_2 = 0.35$ string correlation functions in Fig. \ref{fig11_P} cross at system size $2N \approx {32}$. The functions $g_+(N)$ and $g_-(N)$ are expected to have equal string order $g(\infty) > 0$. The difference $g_+(N) - g_-(N)$ increases to $0.029$ at $2N = 144$ and decreases to $0.026$ at $2N = 192$. Larger systems are required to evaluate the string order. The small gap $\varepsilon_T(1,0.35) = 0.006$ also points to long but finite-ranged spin correlations.


\section{Summary and conclusions\label{sec-VI}}

We have presented spin-$1/2$ string correlation functions and string order in general. The F-AF ladder, Eq. \ref{ssm_eq1}, at specific parameters $J_L$, $J_F$ and $J_A = 1$ reduces to important spin-$1/2$ models with singlet ground states. It has two $N$-spin string correlation functions, $g_1(N)$ and $g_2(N)$, at system size $2N$, $N$ even. Since the ladder is gapped, with $\varepsilon_T(J_L,J_F) > 0$ except in the limit $J_L \to \infty$, the string order is $g_2(\infty) > 0$, $g_1(\infty) = 0$. As shown in Fig. \ref{fig9_p}, the string order $g_2(\infty)$ in the limit $J_F \to \infty$ is equal to $\tilde{g} (\infty)$ of the spin-$1$ HAF, and the limits are approached from below.

The ground state near the origin of the $J_L-J_F$ plane consists of rungs with AF exchange $J_A$ that are weakly coupled by frustrated F exchanges $-J_L$ and $-J_F$ in Fig. \ref{fig1_p}. Short-range spin correlations are indicated by the gaps $\varepsilon_T(J_L,J_F)$ in Fig. \ref{fig2_p}, by string order $g_2(\infty) > 3/4$  and by convergence to the thermodynamic limit at system size $2N = 24$. The regime $J_F > 3$, $J_L \leq 1/2$ has reduced $\varepsilon_T(J_L,J_F)$, finite $g_2(\infty)$ and spin correlations of intermediate range as indicated by the minimum $N^*$ of $g_2(N)$. The regime $J_L > 2$, $J_F \leq 1$ has small $\varepsilon_T(J_L,J_F)$ that vanishes as $1/J_L$ in Fig. \ref{fig4_p}. The range of spin correlations is $N^* \approx 50$ at $J_L = 3$ and increases rapidly with $J_L$. The gapless $J_1-J_2$ model with $J_2 = -J_L$ is the limit $J_L \to \infty$. The model is frustrated when $J_2 > 0$ and illustrates spontaneous dimerization in the dimer phase with finite $\varepsilon_T$ and string order.

String correlation functions of the F-AF ladder directly probe ground-state spin correlations and their range. They afford more nuanced information than the binary choice of finite range in gapped systems and infinite range in gapless systems. The estimated range of spin correlations at $J_L$, $J_F$ and $J_A = 1$ in Eq. \ref{ssm_eq1} is $N^*$, the minimum of $g_2(N)$. The VB interpretation accounts for convergence to string order $g_2(\infty)$ from below and the exponential decrease of $g_1(N)$ for $N \geq N^*$. Ranges up to $N^* \sim 100$ are accessible in DMRG calculations up to system size $2N = 200$.

The spin-1 HAF has one spin-$1$ per unit cell and can be written in terms of two spins-$1/2$ as $s_j = S_{2j-1} + S_{2j}$ with F exchange $-J_F$ in rungs and AF exchange $J/4$ between adjacent rungs. There are now two spins per unit cell and $J_F \to \infty$ excludes singlet-paired rungs. In the VB treatment of finite spin-$1$ HAFs, Eq. \ref{ssm_eq1} was expressed \cite {KChang1989} in terms of spin-$1/2$ operators in a way that gave vanishing matrix elements for diagrams $\lvert q\rangle$ with singlet-paired rungs $2j - 1$, $2j$. F alignment in rungs clearly requires AF exchange and two spins per unit cell in order to have a singlet ground state.

\section{Acknowledgments}
Z.G.S. thanks D. Huse for fruitful discussions. M.K. thanks SERB for financial support through Grant Sanction No. CRG/2020/000754. M.C. thanks DST-INSPIRE for financial support.
\section{Appendix}
We summarize the overlap of singlet VB diagrams and the size dependence of the singlet
sector. In systems of $2N$ spins, singlet diagrams $\lvert q\rangle$ have $N$ lines $(m,n)$ that correspond to normalized singlet-paired spins in Eq. \ref{ssm_eq3} and connect the vertices of the $2N$ polygon without any crossing lines. The overlaps are
\begin{equation} \tag{A1}
{S_{q^\prime q}}=\langle q^\prime \rvert q\rangle = (-2)^{-N+I(q^\prime,q).}
\label{ssm_eq17}
\end{equation} 
$I{(q^\prime,q)}$ is the number of disconnected lines called islands by Pauling when the diagrams are
superimposed. The superposition of any diagram with itself generates $N$ islands of doubled lines
$(m,n)$ and unit overlap. The other extreme, illustrated by $\langle K1\rvert K2\rangle =(-2)^{-N+1}$, is a single island for diagrams without any $(m,n)$ in common. $I{(q^\prime,q)}$ is the number of shared lines $(m,n)$ plus the number of islands with lines connecting vertices at unshared $(m,n)$. The Kekul\'e diagrams have no shared $(m,n)$; their overlap of any $\lvert q \rangle$ satisfies the relation,
\begin{equation} \tag{A2}
{I(q,K1)+I(q,K2)}=N+1
\label{ssm_eq18}
\end{equation}
Overlap magnitudes are necessarily larger with one of the Kekul\'e diagrams when $N$ is even, and overlap magnitude uniquely relates half of the diagrams to $\lvert K1 \rangle$, the other half to $\lvert K2\rangle$. The
Kekul\'e diagrams are orthogonal in the thermodynamic limit, as are diagrams that differ from either by a finite number of lines $(m,n)$.

All eigenfunctions $\lvert q\rangle$ of the string operator $\hat{g_1}(N)$ in Eq. \ref{ssm_eq13} have $1\leq m,n \leq N$. All other $\lvert q\rangle$ have one or more pairs of bridging lines $(m,n)$ with only one end in the string $1$ to $N$. Then $\hat{g_1}(N)\lvert q\rangle$  generates triplet-paired spins $(m,n)_T = ({\alpha_m}{\beta_n} + {\beta_m}{\alpha_n})/\sqrt{2}$ at all bridging lines.
For example, $\hat{g_1}(N)\lvert K2\rangle$ generates diagram $\lvert K^\prime \rangle$ with unchanged $(m,n)$ except for two bridging lines that become $(2N,1)_T$ and $(N,N+1)_T$. The overlap of diagrams with triplets is zero unless the triplets are in the same island, in which case $S_{q q^\prime}$ is Eq. (\ref{ssm_eq17}).  We have  $\langle K^\prime\rvert K2\rangle= 0$ due to spin 
orthogonality and $\langle K^\prime\rvert K1\rangle= (–2)^{–N+1}$ since both triplets are in the same island.

The dimensions of the VB basis have long been known. The number of singlet diagrams in systems of $2N$ spins is
\begin{equation} \tag{A3}
{R_0(2N)}=\frac {(2N)!} {N!(N+1)!}.  \hspace{0.8cm} 
\label{ssm_eq19}
\end{equation}
The string operator for $N$ spins has $R_0(N)$ eigenfunctions $\lvert q\rangle$ with $N/2$ lines in the string. The degeneracy of each is $R_0(N)$ since $\lvert q \rangle$ also has $N/2$ lines with $(m,n)$ not in the string. The
ratio of eigenstates to the total number of singlets is, using Stirling’s approximation,
\begin{equation} \tag{A4}
\frac {{R_0(N)}^2} {R_0(2N)}\approx\frac {8e(N + 1)^{N+3/2}} {\sqrt{\pi}(N+2)^{N+3}}.   \hspace{0.8cm} 
\label{ssm_eq20}
\end{equation}

\bibliography{reference}
\end{document}